\documentclass[preprint,showpacs,preprintnumbers,amsmath,amssymb,prc]{revtex4}

\usepackage{epsfig}
\usepackage{graphicx}
\usepackage{dcolumn}
\usepackage{bm}

\newcommand{\mc}{\multicolumn}
\def\v#1{\mbox{\boldmath$#1$}}

\begin{document}

\title{Microscopic approach to the proton asymmetry in the
non--mesonic weak decay of $\Lambda$--hypernuclei}

\author{E.~Bauer$^1$, G.~Garbarino$^2$, A. Parre\~no$^3$ and A. Ramos$^3$}

\affiliation{$^1$Departamento de F\'{\i}sica, Universidad Nacional de
La Plata, C.C. 67, and IFLP,
CONICET, La Plata, Argentina}
\affiliation{$^2$Dipartimento di Fisica Teorica, Universit\`a di Torino and INFN, Sezione
di Torino, I--10125 Torino, Italy}
\affiliation{$^3$Departament d'Estructura i Constituents de la Mat\`{e}ria,
Universitat de Barcelona,  E--08028 Barcelona, Spain}

\date{\today}

\begin{abstract}
The non--mesonic weak decay of polarized $\Lambda$--hypernuclei is studied with a
microscopic diagrammatic formalism in which one-- and two--nucleon induced decay mechanisms,
$\vec{\Lambda} N \to NN$ and $\vec{\Lambda} NN \to NNN$, are considered together
with (and on the same ground of) nucleon final state interactions.
We adopt a nuclear matter formalism extended to finite nuclei via the local density
approximation. Our approach adopts different one--meson--exchange weak
transition potentials, while the strong interaction effects are accounted
for by a Bonn nucleon--nucleon interaction. We also
consider the two--pion--exchange effect in the weak transition potential.
Both the two-nucleon induced decay mechanism and the final state interactions
reduce the magnitude of the asymmetry.
The quantum interference terms considered in the present microscopic approach
give rise to an opposite behavior of the asymmetry with increasing energy cuts
to that observed in models describing the nucleon final state interactions
semi-classically via the intranuclear cascade code.
Our results for the asymmetry parameter in
$^{12}_{\Lambda}$C obtained with different potential models are consistent with
the asymmetry measured at KEK.
\end{abstract}
\pacs{21.80.+a, 25.80.Pw.}

\maketitle

\newpage

\section{Introduction}
\label{introd}

The study of hypernuclear physics provides the
main source of information on the baryon--baryon strangeness--changing
weak interactions. In particular, the non--mesonic weak
decay of $\Lambda$--hypernuclei has shown two challenging
issues presenting some puzzling character \cite{Al02,Pa07}.
First, we must mention the disagreement between theory
and experiment for the ratio $\Gamma_{n}/\Gamma_{p}$ between the
rates for the $\Lambda n \to nn$ and the $\Lambda p \to np$
non--mesonic weak decay processes. Another more recent
problem concerns the asymmetry in the proton emission from the
non--mesonic weak decay of polarized hypernuclei, which is our main concern in
the present contribution.

A $\Lambda$--hypernucleus can be produced with some degree of polarization.
Indeed, the $n(\pi^+,K^+)\Lambda$ reaction has been used \cite{Aj92-00} at
$p_\pi=1.05$ GeV and small $K^+$ laboratory scattering angles to produce
hypernuclear states with a substantial amount of spin--polarization,
preferentially aligned along the line normal to the reaction plane
which identifies the polarization axis.
The dominant decay mechanisms for a polarized $\Lambda$--hypernucleus are
the following neutron-- and proton--induced processes:
\begin{eqnarray}
\label{dw0}
\vec{\Lambda} n & \rightarrow & nn \\
\label{p-ind}
\vec{\Lambda} p & \rightarrow & np.
\end{eqnarray}
It turns out that the number of protons emitted
parallel to the polarization axis is different from the same quantity measured
in the opposite direction. This asymmetric proton emission is a
consequence of the interference between the parity--conserving and the parity--violating terms
in the $\vec{\Lambda} p \to np$ weak transition potential~\cite{ban90}.

Let us denote with $a^{1N}_{\Lambda}$ the intrinsic asymmetry, arising
from the one--nucleon induced ($1N$) decay in Eq.~(\ref{p-ind}).
The one--nucleon induced decays take place within
the nuclear environment and the resulting nucleon pairs can interact
strongly with others nucleons of the medium before any nucleon leaves the nucleus
and is detected. As a result of these final state
interactions (FSI), the asymmetry measured in an experiment, $a^{\rm
M}_{\Lambda}$,
differs from the intrinsic value, $a^{1N}_{\Lambda}$ .
Most of the theoretical models result in a negative and rather
mass--independent intrinsic asymmetry. Instead, data favor a small
$a^{\rm M}_{\Lambda}$, compatible with a vanishing value, for both
$^5_\Lambda$He and $^{12}_\Lambda$C. This shows a clear
disagreement between $a^{1N}_{\Lambda}$ and $a^{\rm M}_{\Lambda}$.

The reason for this disagreement can be twofold. It can be originated by the
weak decay mechanism itself, which might require some improvement and the
consideration of additional two--nucleon induced processes, and it may also be
due to nucleon FSI. Let us start by considering the various mechanisms which
contribute in the evaluation of $a^{1N}_{\Lambda}$. The theoretical models based
on
one--meson--exchange potentials (OME)~\cite{du96,pa97,pa02,bar05,al05,bar07}
and/or direct quark mechanisms~\cite{sa02} predict $a^{1N}_{\Lambda}$
values in the range from $-0.7$ up to $-0.4$.
By using an effective field theory approach, a dominating central,
spin-- and isospin--independent contact term was predicted in~\cite{pa04} which
allowed the authors to reproduce the experimental total and partial non--mesonic decay
widths for $^{5}_{\Lambda}$He, $^{11}_{\Lambda}$B and $^{12}_{\Lambda}$C,
and the asymmetry parameter for $^{5}_{\Lambda}$He. Motivated by
this work, a scalar--isoscalar $\sigma$--meson--exchange was
added to a $(\pi + K)$--exchange weak model also including a direct quark mechanism
\cite{sa05}. Similarly, the $\sigma$--meson was considered together
with a full OME weak potential in~\cite{bar06}. Although the addition
of the $\sigma$--meson may improve the calculation of $a^{1N}_{\Lambda}$,
it turned out to be not enough to reproduce consistently all the decay data
despite the freedom introduced by the unknown coupling constants of the
$\sigma$--meson.
Later, the OME weak potential was supplemented by the exchange of
(uncorrelated and correlated) two--pion pairs \cite{Ch07}. The
two--pion--exchange potential was obtained from a chiral unitary approach in a
study of the nucleon-nucleon interaction \cite{toki} and was adapted to the
weak sector in~\cite{ji01} for a study of the non--mesonic decay rates,
while the calculation of the asymmetry was also carried out in \cite{Ch07}.
The two--pion exchange mechanism turned out to introduce a significant
central, spin-- and isospin--independent $\Lambda N\to nN$ amplitudes and gave rise to a good
reproduction of the entire set of decay rates and asymmetry data for
$^5_\Lambda$He and $^{12}_\Lambda$C.

We now briefly comment on the effect of FSI on the asymmetry parameter.
First of all, it should be noted that from a strictly quantum--mechanical point of view
the only observables in non--mesonic weak decay are the total
non--mesonic decay width, $\Gamma_{\rm NM}$, the spectra of the emitted
nucleons and the asymmetry $a^{\rm M}_{\Lambda}$ \cite{Ba10b}.
It is the action of FSI which prevents the measurement of any of the
non--mesonic partial decay rates and of the intrinsic asymmetry $a^{1N}_{\Lambda}$.
The link between theory and experiment for both $\Gamma_{n}/\Gamma_{p}$ and
$a^{1N}_{\Lambda}$, is not straightforward, since it is strongly dependent on
FSI. For instance, to obtain the
$\Gamma_{n}/\Gamma_{p}$ ratio from experiments, one should proceed to a
deconvolution of the nucleon rescattering effects contained in the measured
nucleon spectra \cite{ga03}, which requires the use of a theoretical
approach for FSI.
For the asymmetry parameter the situation may seem more
direct, as experimental data for $a^{\rm M}_{\Lambda}$ are available.
However, for a direct comparison with experiment, one must calculate
the asymmetry $a^{\rm M}_{\Lambda}$, which also requires the inclusion of FSI
effects.
Only a couple of approaches \cite{al05,Ch07} calculated this observable in an
appropriate way.
However, both these calculations adopted
an hybrid approach consisting in a shell model for describing the weak decay
and a semi--classical intranuclear cascade (INC) model, for simulating FSI.
The only kind of FSI effects considered within the finite nucleus approach
of \cite{al05,Ch07} are those between the two nucleons emitted in the non--mesonic decay,
which are represented by a wave function describing their relative motion under the
influence of a suitable $NN$ interaction ~\cite{pa02}.
Although a discrepancy with data still remains for proton emission spectra
\cite{ba11,Ba10}, one can safely assert that a formalism which takes care of FSI
leads to a good agreement between theory and experiment concerning
$\Gamma_{n}/\Gamma_{p}$ and $a^{\rm M}_{\Lambda}$.

In the present contribution we evaluate the
asymmetry $a^{\rm M}_{\Lambda}$ employing an alternative approach to
the hybrid one of~\cite{al05,Ch07}.
In our microscopic diagrammatic approach, which was developed
in~\cite{ba07,ba07b,ba11},
both the weak decay and the nucleon FSI are part of the same quantum--mechanical
problem and are thus described in a unified way. Therefore, the present formalism has a
self--consistency that is not present in previous approaches.
The calculation is first performed in nuclear matter and then extended to finite
hypernuclei by means of a local density approximation.
Clearly, our nuclear matter wave functions are less realistic than the shell model ones.
However, FSI are relevant and our quantum--mechanical approach describes them more reliably
than the INC. In~\cite{ba11} we showed that quantum interference terms in the FSI are
very important in the calculations of the observable spectra for the emitted nucleons.
Moreover, in the same work we have called attention on the fact that pure
(i.e., non--quantum interference terms) FSI terms and two--nucleon induced ($2N$) decay
contributions originates from two different time--orderings of the same
Feynman diagrams at second order in the weak transition potential.

Another contribution of the present work is the consideration for the
first time of the $2N$ decays, $\vec{\Lambda} NN \to nNN$,
in a calculation of the asymmetry. We will see that, although these
contributions represent almost 30\% of the decay width, they affect the
asymmetry in a very moderate way.

The work is organized as follows. In Section~\ref{intro} we discuss
general aspects of the asymmetry in the proton emission from
the decay of polarized $\Lambda$--hypernuclei. A formal derivation
of the expressions needed to evaluate the asymmetry parameter is
done in Section~\ref{asym_np} when only $1N$ decays are included, resulting
in the intrinsic asymmetry $a^{1N}_{\Lambda}$,
and in Section~\ref{2N_fsi} when $2N$ decays and FSI are taken into account,
resulting in an approximation for the observable asymmetry $a^{\rm M}_{\Lambda}$.
Numerical results are presented and discussed in Section~\ref{numerics} and
finally, our conclusions are given in
Section~\ref{conclusions}.

\section{General considerations on the asymmetry parameter}
\label{intro}
Spin--polarization observables for baryon--baryon interactions
are important quantities which supply additional information
to the more usual total cross sections or decay rates
and thus facilitate the reconstruction of the interaction amplitudes
from experimental data. For nucleon--nucleon elastic scattering, a
complete study of spin--polarization observables is given, for instance,
in~\cite{By78}. The formal derivation of the asymmetry parameter for the
non--mesonic weak decay of $\Lambda$--hypernuclei is instead provided by \cite{Ra92}.
Here we follow a less conventional analysis in order to remark some
conceptual issues.

Let us denote with $\theta$ the angle between the momentum of the
outgoing proton in the $\vec \Lambda p\to np$ weak process and the
polarization axis of the hypernucleus. The number of emitted protons as a function
of $\theta$ can be written as:
\begin{equation}
\label{Np0}
N_{p}(\theta) = N_{p, \, tot} (1+{\cal{A}}_{y}(\theta))/\pi\, ,
\end{equation}
where $N_{p, \, tot}$ is the total number of emitted protons in the
decay of the polarized $\Lambda$--hypernucleus, while the function
${\cal{A}}_{y}(\theta)$ introduces an asymmetry in the distribution.
By construction, it is evident that:
\begin{equation}
\label{int1}
\int^{\pi}_{0} d \theta \; N_{p}(\theta) = N_{p, \, tot}\, ,
\end{equation}
therefore,
\begin{equation}
\label{int2}
\int^{\pi}_{0} d \theta \; {\cal{A}}_{y}(\theta) = 0\, .
\end{equation}
Eq.~(\ref{int2}) allows one to express ${\cal{A}}_{y}(\theta)$ as a series of
odd powers of $\cos\theta$. By keeping the first term in the
series expansion one has:
\begin{equation}
\label{fun_asym}
{\cal{A}}_{y}(\theta) \cong C \, \cos\theta\, .
\end{equation}
This expression is exact for the scattering of two elementary particles,
as in the present hadronic description of the $\vec \Lambda p \to np$ weak decay.

It is reasonable to write the constant $C$ as the product of the
polarization of the hypernucleus $P_{y}$ times a remaining constant, as follows:
\begin{equation}
\label{const_C}
C \equiv P_{y} \, A_{y}\, ,
\end{equation}
where the $A_{y}$ is the hypernuclear asymmetry parameter.
Being the $\Lambda$ non--mesonic decay in a nucleus a complex process,
it is evident that also the two--body induced decays
$\vec{\Lambda} np  \rightarrow  nnp$ and $\vec{\Lambda} pp  \rightarrow  npp$
as well as FSI terms contribute to the observable proton number
$N_p(\theta)$ of Eq.~(\ref{Np0}).

If we restrict to the number of protons originated from the elemntary
$\vec{\Lambda} p  \rightarrow  np$ process, the shell model weak--coupling
scheme allows us to make the following replacement:
\begin{equation}
\label{const_C2}
P_{y} \, A_{y} \rightarrow p_{\Lambda}\, a^{1N}_{\Lambda}\, ,
\end{equation}
by introducing the $\Lambda$ polarization $p_{\Lambda}$
and the intrinsic $\Lambda$ asymmetry $a^{1N}_{\Lambda}$~
\footnotemark{\footnotetext{
In the shell model weak--coupling limit it is easy to obtain
$p_{\Lambda}  =  - J/(J+1) \, P_{y}$ for $J=J_{C}-1/2$ and
$p_{\Lambda}  =  P_{y}$ for $J=J_{C}+ 1/2$, where $J$ $(J_{C})$ is the total
angular momentum of the hypernucleus (core nucleus). For
nuclear matter we have $J_{C}=0$ and then $p_{\Lambda}  =  P_{y}$.}}.
With these definitions, if the weak--coupling limit provides a reliable description
of the hypernucleus, the intrinsic asymmetry $a^{1N}_{\Lambda}$ has the same value
for any hypernuclear species.

We can thus rewrite Eq.~(\ref{Np0}) as follows:
\begin{equation}
\label{Npwd}
N^{1N}_{p}(\theta) = N^{1N}_{p, \, tot} (1+p_{\Lambda} a^{1N}_{\Lambda}  \;
\cos\theta)\, ,
\end{equation}
where the index $1N$ refers to the fact that we are considering only
the one--nucleon induced decay $\vec \Lambda p \to np$.
From this expression, the intrinsic asymmetry is obtained as:
\begin{equation}
\label{asimt}
a^{1N}_{\Lambda} = \frac{1}{p_{\Lambda}}
\, \frac{N^{1N}_{p}(0^{0}) - N^{1N}_{p}(180^{0}) }
{N^{1N}_{p}(0^{0}) + N^{1N}_{p}(180^{0})}\, .
\end{equation}

Once we consider the two--body induced decay process and FSI as well, the
number of emitted protons takes the following form:
\begin{equation}
\label{nptot}
N_{p}(\theta) \equiv N^{1N}_{p}(\theta) + N^{2N+FSI}_{p}(\theta)\, .
\end{equation}
As long as $N^{2N+FSI}_{p}(\theta)$ has a linear dependence on
$\cos\theta$, it is possible to define an observable asymmetry
parameter, $a_{\Lambda}^{1N+2N+FSI}$, given by a relation which is analogous to the one
in Eq.~(\ref{asimt}), which can be compared with the experimental data
for the asymmetry $a_{\Lambda}^{\rm M}$.

\section{Formal derivation of the intrinsic asymmetry}
\label{asym_np}

For computational purposes, we may assume that the hypernucleus is completely
polarized. The intrinsic asymmetry is then
given by Eq.(\ref{asimt}) with $p_\Lambda=P_{y}=1$.

We now focus on the evaluation of the $N^{1N}_{p}(\theta)$ spectrum.
For our practical purpose, we can suppose that the polarized $\Lambda$ has
its spin aligned with the polarization axis
(which thus coincides with the quantization axis).
The evaluation of $N^{1N}_{p}(\theta)$ is rather similar
to the evaluation of the proton kinetic energy spectrum $N_{p}(T_p)$
described in~\cite{ba07,ba07b,ba11}, except for two points: $i)$ the angle
$\theta$ replaces the proton kinetic energy $T_p$ as variable and $ii)$
we no longer sum over the two spin projections of the $\Lambda$, but retain
only the up component. This second point requires
a new evaluation of the spin--summation.
To build up an analytical expression for $N^{1N}_{p}(\theta)$,
let us first express it in terms of the more familiar decay widths as follows:
\begin{equation}
\label{npspec0}
N^{1N}_{p}(\theta) = \bar{\Gamma}_{p}(\theta)\, ,
\end{equation}
with $\bar{\Gamma}_{p}(\theta) \equiv \Gamma_{p}(\theta)/\Gamma_{\rm NM}$,
where $\Gamma_{\rm NM}$ is the total non--mesonic weak decay rate and $\Gamma_{p}(\theta)$
is the proton induced decay rate as a function of
$\theta$~\footnotemark{\footnotetext{Note that the proton--induced
decay rates is obtained as $\Gamma_p=\int_{0^\circ}^{180^\circ}d \theta\, \Gamma_p(\theta)$.}}.
With these definitions, the $N^{1N}_{p}(\theta)$ spectrum is normalized per
non--mesonic weak decay.

Before we give explicit expressions for the $\theta$--dependent proton spectrum, it is
convenient to introduce first the weak transition potential:
\begin{equation}
\label{intlnnn}
V^{\Lambda N\to NN} (q) = \sum_{\tau=0,1} {\cal O}_{\tau}
{\cal V}_{\tau}^{\Lambda N \to NN} (q)~,
\end{equation}
where the isospin dependence is given by
\begin{eqnarray}
\label{isos} {\cal O}_{\tau} =
\left\{
\begin{array}{c}~1~~~~~\mbox{for}~~\tau=0\\
  \v{\tau}_1 \cdot \v{\tau}_2~\mbox{for}~~\tau=1~.
\end{array}\right.
\end{eqnarray}
The values $0$ and $1$ for $\tau$ refer to the isoscalar
and isovector parts of the interactions, respectively.
The spin and momentum dependence of the weak transition potential is
given by the function:
\begin{eqnarray}
\label{intln}
{\cal V}_{\tau}^{\Lambda N \to NN} (q) &
= &  (G_F m_{\pi}^2)  \; \{
S_{\tau}(q)  \; \v{\sigma}_1 \cdot \v{\hat{q}} +
S'_{\tau}(q)  \; \v{\sigma}_2 \cdot \v{\hat{q}} +
P_{C, \tau}(q) \\
& &
+ P_{L, \tau}(q)
\v{\sigma}_1 \cdot \v{\hat{q}} \; \v{\sigma}_2 \cdot
\v{\hat{q}}  +
P_{T, \tau}(q)
(\v{\sigma}_1 \times \v{\hat{q}}) \cdot (\v{\sigma}_2 \times
\v{\hat{q}}) \nonumber \\
& &
+i S_{V, \tau}(q)
\v{(\sigma}_1 \times \v{\sigma}_2) \cdot
\v{\hat{q}} \}~, \nonumber
\end{eqnarray}
where the index 1 (2) refers to the strong (weak) vertex.
The functions $S_{\tau}(q)$, $S'_{\tau}(q)$,
$P_{C, \tau}(q)$, $P_{L,\tau}(q)$,
$P_{T, \tau}(q)$ and $S_{V, \tau}(q)$,
which include short range correlations,
can be adjusted to reproduce any weak transition potential.
Explicit expressions can be found in~\cite{ba03}.
The $S's$ ($P's$) functions are the parity--violating (parity--conserving)
contributions of the weak transition potential.

\begin{figure}[h]
\centerline{\includegraphics[scale=0.63]{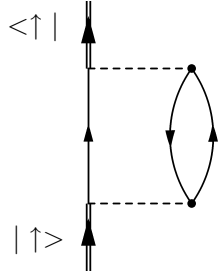}}
\caption{Direct Goldstone diagram corresponding to the square of the
$\vec{\Lambda} p  \rightarrow  np$ decay amplitude.}
\label{spinup}
\end{figure}

In Fig.~\ref{spinup} we show the Goldstone diagram which
has to be evaluated in the calculation of $N^{1N}_{p}(\theta)$.
The spin summation for this diagram is performed for
all particles except for the $\Lambda$, which is assumed to
have spin up. This summation reads:
\begin{eqnarray}
\label{tdir}
{\cal S}^{dir, \, \uparrow}_{\tau \tau'}(q) & = &  2 \; \{
S_{\tau}(q) S_{\tau'}(q) + S'_{\tau}(q) S'_{\tau'}(q)
 + P_{L, \tau}(q)  P_{L, \tau'}(q)  +  P_{C, \tau}(q) P_{C,
\tau'}(q) \\
 & & + 2  \, P_{T, \tau}(q)  P_{T, \tau'}(q) + 2 \, S_{V,
\tau}(q) S_{V, \tau'}(q) \nonumber \\
&& - 2 \, [ S_{\tau}(q) P_{C,\tau'}(q) + S_{\tau'}(q) P_{C,\tau}(q) +
S'_{\tau}(q) P_{L, \tau'}(q) + S'_{\tau'}(q) P_{L, \tau}(q) \nonumber \\
 & & + 2  \,S_{V, \tau}(q) P_{T, \tau'}(q)  + 2  \,S_{V, \tau'}(q) P_{T, \tau}(q)] \, \hat{q_{z}}
\}\, . \nonumber
\end{eqnarray}
It is instructive to note that in the summation over the $\Lambda$ spin
projection,
\begin{eqnarray}
\label{tdirsum}
{\cal S}^{dir, \, \uparrow}_{\tau \tau'}(q)  +
{\cal S}^{dir, \, \downarrow}_{\tau \tau'}(q)
& = &  4 \; \{
S_{\tau}(q) S_{\tau'}(q) + S'_{\tau}(q) S'_{\tau'}(q)
 + P_{L, \tau}(q)  P_{L, \tau'}(q)  +  P_{C, \tau}(q) P_{C,
\tau'}(q) \nonumber \\
 & & + 2  \, P_{T, \tau}(q)  P_{T, \tau'}(q) + 2 \, S_{V,
\tau}(q) S_{V, \tau'}(q) \}\, ,
\end{eqnarray}
the terms between square brackets in~Eq.~(\ref{tdir}) are no longer present.
These terms
are responsible for the asymmetry parameter and are clearly due to interferences
between parity--violating and parity--conserving contributions of the weak
transition potential in Eq.~(\ref{intln}).

\begin{figure}[h]
\centerline{\includegraphics[scale=0.63]{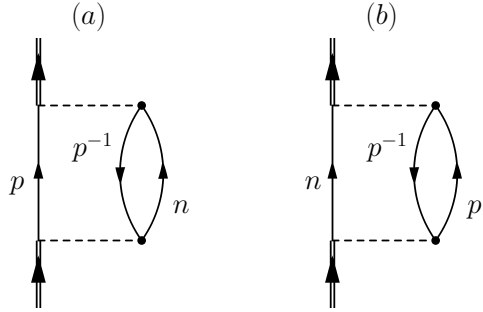}}
\caption{The two isospin contributions to the direct Goldstone
diagram for the $\vec \Lambda p \to np$ decay.
$(a)$ is the charge--exchange contribution, which
is not vanishing only for $\tau=\tau'=1$, while $(b)$ is the charge--conserving one.
The double arrows represent the $\Lambda$ and carry a momentum $\v{k}$,
while dashed lines represent the weak transition potential with momentum $\v{q}$.
The momentum assigned to each intermediate nucleon line is:
$\v{k}-\v{q}$ for the left-most nucleon line ($p$ in (a) or $n$ in (b)), $\v{h}$
for $p^{-1}$ and $\v{h}+\v{q}$ for the right-most nucleon line ($n$ in (a) or
$p$ in (b)).}
\label{exccon}
\end{figure}

Following~\cite{ba07,ba07b}, we introduce now a partial, isospin--dependent
decay width, $\Gamma^{(i)}_{\tau \, \tau'}(\v{k},k_F,\theta)$,
where $\v{k}$ is the momentum of the $\Lambda$ and $k_F$ the Fermi momentum of nuclear matter.
This is done for the two isospin channels, $\tau$, $\tau'=0,1$, contributing to the spectra.
In Fig.~\ref{exccon} we depict the charge--exchange and
charge--conserving contributions. The distinction between the two terms is important
in the evaluation of $N^{1N}_p(\theta)$ as the kinematics of the
proton attached to the weak vertex is different from the one outgoing from the strong vertex.
The partial, isospin--dependent decay widths for the two terms of Fig.~\ref{exccon}
are:
\begin{eqnarray}
\label{gamdira}
\Gamma^{(a)}_{1,1}(\v{k},k_F,\theta)  & =
& (G_F m_{\pi}^2)^2 \frac{1}{(2 \pi)^5} \int  \int d
\v{q} d \v{h} \; {\cal S}^{dir, \, \uparrow}_{11}(q)  \;
\theta(q_0) \theta(|\v{k-q}|-k_F)  \\
& & \times \theta(|\v{h}+\v{q}|-k_F) \theta(k_F-
|\v{h}|)  \;
\delta(q_0 - (E_N(\v{h}+\v{q}) - E_N(\v{h}))) \nonumber \\
&& \times \delta(\cos\theta-(\v{k-q})_{z}/|\v{k-q}|) \, , \nonumber
\end{eqnarray}
and
\begin{eqnarray}
\label{gamdirb}
\Gamma^{(b)}_{\tau \, \tau'}(\v{k},k_F,\theta)  & =
& (G_F m_{\pi}^2)^2 \frac{1}{(2 \pi)^5} \int  \int d
\v{q} d \v{h} \; {\cal S}^{dir, \, \uparrow}_{\tau \tau'}(q)  \;
\theta(q_0) \theta(|\v{k-q}|-k_F) \\
& & \times \theta(|\v{h}+\v{q}|-k_F) \theta(k_F-
|\v{h}|)  \;
\delta(q_0 - (E_N(\v{h}+\v{q}) - E_N(\v{h}))) \nonumber \\
&& \times \delta(\cos\theta-(\v{h+q})_{z}/|\v{h+q}|) \, , \nonumber
\end{eqnarray}
where the kinematics is explained in Fig.~\ref{exccon}. Label $(a)$ refers
to the charge--exchange contribution (proton attached to the $\Lambda$ vertex)
and label $(b)$ represents the charge--conserving term (proton attached in
the
strong vertex). In previous equations one has $q_0=k_0 - E_N(\v{k-q}) - V_N$,
$k_0$ being the total energy of the $\Lambda$,
$E_N$ the nucleon total free energy and $V_N$ the nucleon binding energy.
After performing the isospin summation we obtain:
\begin{equation}
\label{isospin}
\Gamma_{p} = 4 \Gamma^{(a)}_{1,1} + \Gamma^{(b)}_{1,1}
+ \Gamma^{(b)}_{0,0} - \Gamma^{(b)}_{0,1}-\Gamma^{(b)}_{1,0}\, ,
\end{equation}
where the $(\v{k},k_F,\theta)$--dependence of all functions
is omitted for simplicity. Finally, the
decay rates for a finite hypernucleus are obtained by the local density approximation,
i.e., after averaging the above partial width over the $\Lambda$
momentum distribution in the considered hypernucleus,
$|\widetilde{\psi}_{\Lambda}(\v{k})|^2$, and over the local Fermi momentum,
$k_{F}(r) = \{3 \pi^{2} \rho(r)/2\}^{1/3}$,
$\rho(r)$ being the density profile of the nuclear core.
One thus has:
\begin{equation}
\label{decwpar3}
\Gamma_{p}(\theta) = \int d \v{k} \, |\widetilde{\psi}_{\Lambda}(\v{k})|^2
\int d \v{r} \, |\psi_{\Lambda}(\v{r})|^2
\Gamma_{p}(\v{k},k_{F}(r),\theta)~,
\end{equation}
where $\psi_{\Lambda}(\v{r})$ is the Fourier transform of
$\widetilde{\psi}_{\Lambda}(\v{k})$. The $\Lambda$ total energy
is given by $k_{0}=m_\Lambda+\v{k}^2/(2 m_\Lambda)+V_{\Lambda}$,
where $V_\Lambda$ is a binding potential.

Finally, by inserting the quantities
$N^{1N}_{p}(\theta)=\bar{\Gamma}_{p}(\theta)=\Gamma_{p}(\theta)/\Gamma_{\rm NM}$
for $0^{0}$ and $180^{0}$ in Eq.~(\ref{asimt}) with $p_\Lambda=1$, the
intrinsic asymmetry
$a^{1N}_\Lambda$ is obtained.

\section{Effect of the strong interaction on the asymmetry}
\label{2N_fsi}
The evaluation of the asymmetry $a^{1N+2N+FSI}_{\Lambda}$,
which includes the effects of both $2N$ and FSI--induced decay processes,
is an involved task and, up to now, analytical expressions were given only for the
intrinsic asymmetry $a^{1N}_{\Lambda}$, while numerical calculations were performed for
$a^{1N+FSI}_{\Lambda}$ by using the aforementioned hybrid approach incorporating the
INC \cite{al05,Ch07}.
In this section we present for the first time analytical expressions
for $a^{1N+2N+FSI}_{\Lambda}$.

\begin{figure}[h]
\centerline{\includegraphics[scale=0.73]{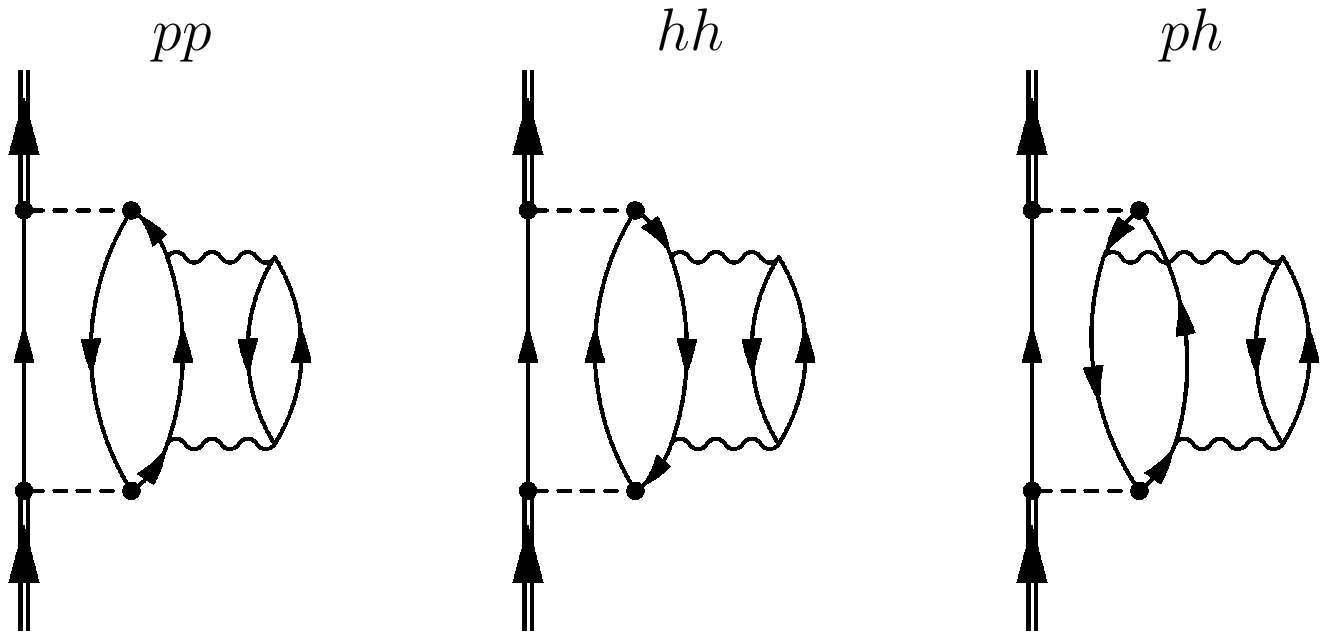}}
\caption{The set of Feynman diagrams considered in this work for the in--medium $\Lambda$
self--energy. The different time--ordering terms of these diagrams contribute to $2N$ and
FSI--induced decays.}
\label{2Nfsi}
\end{figure}

We follow similar steps as in the last section in order to derive $N^{2N+FSI}_{p}(\theta)$,
which provides the total proton spectrum $N^{1N+2N+FSI}_{p}(\theta) = N^{1N}_{p}(\theta) +
N^{2N+FSI}_{p}(\theta)$. This is done by introducing the set of Feynman diagrams
depicted in Fig.~\ref{2Nfsi} to take care of $2N$ decays and FSI effects which result from the
action of the nucleon--nucleon strong interaction involving the nucleons
produced by the weak decay and nucleons of the medium. The choice of the set
of diagrams in Fig.~\ref{2Nfsi} is motivated by previous calculations~\cite{ba07,ba07b,ba11},
which show that these are the dominant contributions in the evaluation of
the nucleon emission spectra. Each Feynman diagram is the sum of a number of time--ordering (i.e,
Goldstone) diagrams. It is in terms of these Goldstone diagrams that one can differentiate among
$1N$, $2N$, pure FSI and quantum interference terms (QIT) between $1N$ or $2N$ and FSI
contributions.
This point is relevant as it shows that, from a quantum--mechanical perspective,
each of the above processes are included in a unitary description.
More details on this point are given in~\cite{ba11}.

Since in the evaluation with Goldstone diagrams the $2N$ decays are separated
contributions from FSI--induced decays (which are divided in pure FSI and QIT terms),
the $2N+FSI$ proton spectrum reads:
\begin{eqnarray}
\label{np1f-0}
N^{2N+FSI}_{p}(\theta)=N^{2N}_{p}(\theta)+N^{FSI}_{p}(\theta)\, ,
\end{eqnarray}
where:
\begin{eqnarray}
\label{np1f}
N^{2N}_{p}(\theta) &=& \bar{\Gamma}_{np}(\theta) + 2 \bar{\Gamma}_{pp}(\theta)~, \\
\label{np1f-2}
N^{FSI}_{p}(\theta) &=& \sum_{i, \, f} N_{f} \, \bar{\Gamma}_{i, f}(\theta)~.
\end{eqnarray}
Here, $\bar{\Gamma} \equiv \Gamma/\Gamma_{\rm NM}$ stands for the decay rate of
a particular decay mode normalized per non-mesonic weak decay. The functions
$\bar \Gamma_{np}$
and $\bar \Gamma_{pp}$ represent the
$\vec \Lambda np \to nnp$ and $\vec \Lambda pp\to npp$ decay processes,
respectively,
while $\bar{\Gamma}_{i, f}$
represent either pure FSI Goldstone diagrams or QIT Goldstone
diagrams, accounting for the quantum interference among $1N$ or $2N$ and
FSI--induced decay processes.
The index $i$ in $\bar{\Gamma}_{i, f}$ is used to label a particular Goldstone
diagram obtained from the Feynman diagrams in Fig.~\ref{2Nfsi},
while $f$ denotes the final physical states of the Goldstone diagram and in the
present case can take the values $f=np$ (cut on $2p1h$ states) and $f=npN$ (cut on $3p2h$
states), with $N=n$, $p$, since we need at least one proton in the final state
to obtain $N^{1N+2N+FSI}_p(\theta)$. Finally, $N_{f}$ is the number of
protons contained in the multinucleon state $f$.

At this point it is necessary to introduce the adopted nucleon--nucleon strong potential:
\begin{equation}
\label{intnn1}
V^{NN} (t) = \sum_{\tau_{N}=0,1}
 {\cal O}_{\tau_{N}}
{\cal V}_{\tau_{N}}^{NN} (t)~,
\end{equation}
where $t$ is the momentum carried by the strong interaction,
${\cal O}_{\tau_{N}}$ is defined in Eq.~(\ref{isos})
and the spin and momentum dependence of the interaction is given by:
\begin{eqnarray}
\label{intnn2}
{\cal V}_{\tau_N}^{N N} (t)
& = & \frac{f_{\pi}^2}{m_{\pi}^2}  \; \{
{\cal V}_{C, \,\tau_{N}}(t) +
{\cal V}_{L, \, \tau_{N}}(t)
\v{\sigma}_1 \cdot \v{\hat{t}} \; \v{\sigma}_2 \cdot
\v{\hat{t}} \\
& & + {\cal V}_{T, \, \tau_{N}}(t)
(\v{\sigma}_1 \times \v{\hat{t}}) \cdot (\v{\sigma}_2 \times
\v{\hat{t}}) \}~, \nonumber
\end{eqnarray}
where the functions ${\cal V}_{C, \,\tau_{N}}(t)$, ${\cal V}_{L, \, \tau_{N}}(t)$ and
${\cal V}_{T, \, \tau_{N}}(t)$ are adjusted to reproduce any strong interaction.

In the calculation of the diagrams in Fig.~\ref{2Nfsi} the
isospin summation is particularly complex as one has to
differentiate the isospin projection of each particle.
We give details on this aspect in the present Section and in
the Appendix. The main features of the momentum dependence of the diagrams
were discussed in~\cite{ba11} and references therein. However, the important point
in the evaluation of the asymmetry is the spin dependence of the diagrams.

We thus start by considering the spin summation for each Goldstone diagram obtained from
Fig.~\ref{2Nfsi}. This sum is performed for all particles except the
$\Lambda$, which again
is assumed to have spin up with respect to the polarization axis.
For the diagrams $pp$ and $hh$ we obtain:
\begin{equation}
\label{spphh}
{\cal S}^{pp \, (hh)}_{\tau \tau'; \tau_{N} {\tau'}_{N}}(q,t)  =   2 \;
{\cal S}^{dir, \, \uparrow}_{\tau \tau'}(q) \;
\{ {\cal V}_{C, \, \tau_{N}}(t) {\cal V}_{C, \, \tau'_{N}}(t)  +
{\cal V}_{L, \, \tau_{N}}(t) {\cal V}_{L, \, \tau'_{N}}(t)
+ 2 \, {\cal V}_{T, \, \tau_{N}}(t) {\cal V}_{T, \, \tau'_{N}}(t)\}\, ,
\end{equation}
where ${\cal S}^{dir, \, \uparrow}_{\tau \tau'}(q)$ is given in Eq.~(\ref{tdir}).
The spin summation is more complex for the $ph$ diagram. It is
convenient to split it in the sum of two terms:
\begin{equation}
\label{spinph}
{\cal S}^{ph}_{\tau \tau'; \tau_{N} {\tau'}_{N}}(q,t) \equiv
{\cal S}^{ph, \,\rm no-asym}_{\tau \tau'; \tau_{N} {\tau'}_{N}}(q,t) +
{\cal S}^{ph, \, \rm asym}_{\tau \tau'; \tau_{N} {\tau'}_{N}}(q,t)\, ,
\end{equation}
where
\begin{eqnarray}
\label{sph}
{\cal S}^{ph, \, \rm no-asym}_{\tau \tau'; \tau_{N} {\tau'}_{N}}(q,t) & = &  4 \; \{
(S_{\tau}(q) S_{\tau'}(q) +  P_{C,\tau}(q) P_{C,\tau'}(q)) \; {\cal W}^{C}_{\tau_{N} {\tau'}_{N}}(t)  \\
&& \; + (S'_{\tau}(q) S'_{\tau'}(q) +  P_{L, \tau}(q) P_{L, \tau'}(q))
\; {\cal W}^{L}_{\tau_{N} {\tau'}_{N}}(t) \nonumber \\
&& \; + 2  (\,S_{V, \tau}(q)  S_{V, \tau'}(q) +  P_{T, \tau}(q) P_{T, \tau'}(q))
\; {\cal W}^{C}_{\tau_{N} {\tau'}_{N}}(t) \}\, , \nonumber
\end{eqnarray}
represents the term which does not contribute to the asymmetry and
\begin{eqnarray}
\label{sphasim}
{\cal S}^{ph, \,\rm asym}_{\tau \tau'; \tau_{N} {\tau'}_{N}}(q,t) & = &  -8 \; \{
(S_{\tau}(q) P_{C,\tau'}(q) + S_{\tau'}(q) P_{C,\tau}(q)) \; {\cal W}^{C}_{\tau_{N} {\tau'}_{N}}(t) \\
&& \; +(S'_{\tau}(q) P_{L, \tau'}(q) + S'_{\tau'}(q) P_{L, \tau}(q))
\; {\cal W}^{L}_{\tau_{N} {\tau'}_{N}}(t) \nonumber \\
&& \;  +2  (\,S_{V, \tau}(q) P_{T, \tau'}(q)  + S_{V, \tau'}(q) P_{T, \tau}(q))
\; {\cal W}^{C}_{\tau_{N} {\tau'}_{N}}(t) \} \, \hat{q_{z}}\, , \nonumber
\end{eqnarray}
is the term responsible for the asymmetry. In these expressions we have
introduced the functions:
\begin{eqnarray}
\label{intdirasim}
{\cal W}^{C}_{\tau_{N} {\tau'}_{N}} & = &
{\cal V}_{C, \, \tau_{N}} {\cal V}_{C, \, \tau'_{N}}  +
{\cal V}_{L, \, \tau_{N}} {\cal V}_{L, \, \tau'_{N}}
+ 2 \, {\cal V}_{T, \, \tau_{N}} {\cal V}_{T, \, \tau'_{N}}\, , \\
{\cal W}^{L}_{\tau_{N} {\tau'}_{N}} & = &
{\cal V}_{C, \, \tau_{N}} {\cal V}_{C, \, \tau'_{N}}  -
{\cal V}_{T, \, \tau_{N}} {\cal V}_{T, \, \tau'_{N}}
+ (-1+2 \hat{q} \cdot \hat{t}) \, ({\cal V}_{L, \, \tau_{N}} {\cal V}_{L, \, \tau'_{N}}
-{\cal V}_{T, \, \tau_{N}} {\cal V}_{T, \, \tau'_{N}})\, , \nonumber \\
{\cal W}^{T}_{\tau_{N} {\tau'}_{N}} & = &
{\cal V}_{C, \, \tau_{N}} {\cal V}_{C, \, \tau'_{N}}  +
{\cal V}_{T, \, \tau_{N}} {\cal V}_{T, \, \tau'_{N}}
+ \hat{q} \cdot \hat{t} \, ({\cal V}_{L, \, \tau_{N}} {\cal V}_{L, \, \tau'_{N}}
-{\cal V}_{T, \, \tau_{N}} {\cal V}_{T, \, \tau'_{N}})\, , \nonumber
\end{eqnarray}
representing the effect of the strong interaction. For simplicity, the
$t$--dependence in both the ${\cal W}$'s and ${\cal V}$'s
has been omitted in these expressions. Although Eq.~(\ref{sphasim})
is a more complicated expression than the ones in Eqs.~(\ref{tdir}) and
(\ref{spphh}), again the asymmetry is originated from the
interference between parity--violating ($S$'s) and parity--conserving ($P$'s)
terms of the weak transition potential.

\begin{figure}[h]
\centerline{\includegraphics[scale=0.67]{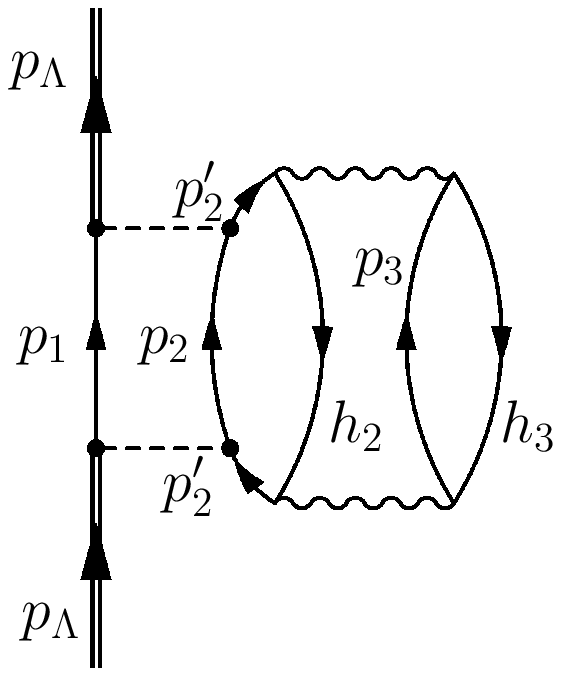}}
\caption{Time--ordering term from Feynman diagram
$hh$ in Fig.~\ref{2Nfsi} contributing to $2N$ decays. The momentum carried by
the weak transition potential (strong interaction) is $\v{q}$ ($\v{t}$). The
momentum of each particle is $\v{k}$ for $p_{\Lambda}$,
$\v{k}-\v{q}$ for $p_1$, $\v{k}-\v{t}+\v{q}$ for $p_2$,
$\v{h}+\v{t}$ for $p'_2$,
$\v{h}$ for $h_2$, $\v{h'}+\v{t}$ for $p_3$ and
$\v{h'}$ for $h_3$.}
\label{2Nd}
\end{figure}

The next step is to implement the momentum and isospin summation for
each Goldstone diagram. In this Section we choose the Goldstone diagram in
Fig.~\ref{2Nd} as a representative example for this evaluation
and we leave to the Appendix the remaining contributions.
This diagram is a particular time--ordering contribution stemming from the
$hh$ Feynman diagram in Fig.~\ref{2Nfsi}.
It contributes to the two--nucleon induced decay mechanisms
$\vec{\Lambda} np  \rightarrow  nnp$ and $\vec{\Lambda} pp \rightarrow npp$
with one and two protons in the final states, respectively.

Protons in the final state can be in any of the nucleon lines labelled by $p_1$,
$p_2$ and $p_3$ in Fig.~\ref{2Nd}. To deal with this matter, it is again
convenient to introduce some partial, isospin dependent decay widths.
For the $\vec{\Lambda} np  \rightarrow  nnp$
Goldstone diagram of Fig.~\ref{2Nd} we define the following rates:
\begin{eqnarray}
\label{gamhhp1}
\Gamma^{hh, \, p_1}_{\tau \tau'; \tau_{N} {\tau'}_{N}}(\v{k},k_F,\theta) & =
& \frac{(G_F m_{\pi}^2)^{2}}{(2 \pi)^{5}}
\left(\frac{f_{\pi}^2}{4 \pi}\right)^{2}  \frac{1}{m_{\pi}^4}
\frac{2}{(2 \pi)^2} \;
\int d  \v{q}
\int d  \v{t} \int d \v{h} \int d \v{h'} \,
\theta(q_{0}) \\
&& \times \theta(|\v{k}-\v{q}| - k_F)
\theta(|\v{h}-\v{t}| - k_F) \, \theta(|\v{h}-\v{t}+\v{q}| - k_F)
\theta(k_F - |\v{h}|) \nonumber \\
&&
\times \theta(|\v{h'}+\v{t}| - k_F) \,
\theta(k_F - |\v{h'}|) \; \delta(\cos\theta-(\v{k-q})_{z}/|\v{k-q}|)
\nonumber \\
&&
\times \delta(q_0 - (E_N(\v{h}-\v{t}+\v{q}) - E_N( \v{h})+E_N(\v{h'}+\v{t})-E_N(\v{h'}))
\nonumber \\
& &
\times \frac{{\cal S}^{hh}_{\tau \tau'; \tau_{N} {\tau'}_{N}}(q,t) }
{(E_N(\v{h}-\v{t})-E_N(\v{h})+E_N(\v{h'}+\v{t})-E_N(\v{h'}))^{2}}\, ,\nonumber
\end{eqnarray}
\begin{eqnarray}
\Gamma^{hh, \, p_2}_{\tau \tau'; \tau_{N} {\tau'}_{N}}(\v{k},k_F,\theta) & =
& \Gamma^{hh, \, p_1}_{\tau \tau'; \tau_{N} {\tau'}_{N}}(\v{k},k_F,\theta)
| \\
&&\delta(\cos\theta-(\v{k-q})_{z}/|\v{k-q}|) \rightarrow
\delta(\cos\theta-(\v{h-t+q})_{z}/|\v{h-t+q}|)\, , \nonumber \\
&& \nonumber \\
\Gamma^{hh, \, p_3}_{\tau \tau'; \tau_{N} {\tau'}_{N}}(\v{k},k_F,\theta) & =
& \Gamma^{hh, \, p_1}_{\tau \tau'; \tau_{N} {\tau'}_{N}}(\v{k},k_F,\theta)
| \\
&&\delta(\cos\theta-(\v{k-q})_{z}/|\v{k-q}|) \rightarrow
\delta(\cos\theta-(\v{h'+t})_{z}/|\v{h'+t}|)\, , \nonumber
\end{eqnarray}
each one for the final proton occupying the nucleon line $p_1$, $p_2$ or $p_3$
in the diagram, respectively.
In a similar way, for the reaction $\vec{\Lambda} pp  \rightarrow  npp$, we have:
\begin{eqnarray}
\label{gamhhp1p3}
\Gamma^{hh, \, p_1, \, p_3}_{\tau \tau'; \tau_{N} {\tau'}_{N}}(\v{k},k_F,\theta) & =
& \frac{(G_F m_{\pi}^2)^{2}}{(2 \pi)^{5}}
\left(\frac{f_{\pi}^2}{4 \pi}\right)^{2}  \frac{1}{m_{\pi}^4}
\frac{2}{(2 \pi)^2} \;
\int d  \v{q}
\int d  \v{t} \int d \v{h} \int d \v{h'} \, \theta(q_{0}) \\
&& \times \theta(|\v{k}-\v{q}| - k_F)
\theta(|\v{h}-\v{t}| - k_F) \, \theta(|\v{h}-\v{t}+\v{q}| - k_F)
\theta(k_F - |\v{h}|)
\nonumber \\
&&
\times \theta(|\v{h'}+\v{t}| - k_F) \,
\theta(k_F - |\v{h'}|) \nonumber \\
&&\times (\delta(\cos\theta-(\v{k-q})_{z}/|\v{k-q}|)+
\delta(\cos\theta-(\v{h'+t})_{z}/|\v{h'+t}|))/2
\nonumber \\
&&
\times \delta(q_0 - (E_N(\v{h}-\v{t}+\v{q}) - E_N( \v{h})+E_N(\v{h'}+\v{t})-E_N(\v{h'}))
\nonumber \\
& &
\times \frac{{\cal S}^{hh}_{\tau \tau'; \tau_{N} {\tau'}_{N}}(q,t) }
{(E_N(\v{h}-\v{t})-E_N(\v{h})+E_N(\v{h'}+\v{t})-E_N(\v{h'}))^{2}}\, ,\nonumber \\
&& \nonumber \\
\Gamma^{hh, \, p_2, \, p_3}_{\tau \tau'; \tau_{N} {\tau'}_{N}}(\v{k},k_F,\theta) & =
& \Gamma^{hh, \, p_1, \, p_3}_{\tau \tau'; \tau_{N} {\tau'}_{N}}(\v{k},k_F,\theta)| \\
&&\delta(\cos\theta-(\v{k-q})_{z}/|\v{k-q}|) \rightarrow
\delta(\cos\theta-(\v{h-t+q})_{z}/|\v{h-t+q}|)\, , \nonumber
\end{eqnarray}
where the sum of the two delta functions in $\cos\theta$ in Eq.(\ref{gamhhp1p3})
is divided by two in order to retain this multiplicative factor in front of
$\bar{\Gamma}_{pp}(\theta)$ in Eq.~(\ref{np1f}).
Note that charge conservation does not allow particles $p_1$ and $p_2$ to be two protons
simultaneously.

The next step is to implement the isospin summation. For the
$\vec{\Lambda} np  \rightarrow  nnp$ decay we obtain:
\begin{eqnarray}
\label{gamma2np}
\Gamma^{hh, \, p_1}_{np} & = & 4 (
5 {\Gamma}^{hh, \,p_1}_{11,11} + {\Gamma}^{hh, \,p_1}_{00,00} -
2 {\Gamma}^{hh, \,p_1}_{11,00})\, , \\
\Gamma^{hh, \,p_2}_{np}  & = &
5 {\Gamma}^{hh, \,p_2}_{11,11} + {\Gamma}^{hh, \,p_2}_{00,00} - 10 {\Gamma}^{hh, \,p_2}_{01,11} +
5 {\Gamma}^{hh, \,p_2}_{00,11} \nonumber \\
&& + {\Gamma}^{hh, \,p_2}_{11,00} -2 {\Gamma}^{hh, \,p_2}_{01,00} - 2 {\Gamma}^{hh, \,p_2}_{11,01} -
2 {\Gamma}^{hh, \,p_2}_{11,01} \nonumber \\
&& - 2 {\Gamma}^{hh, \,p_2}_{00,01} + 4 {\Gamma}^{hh, \,p_2}_{01,01}\, ,  \nonumber \\
\Gamma^{hh, \,p_3}_{np}  & = &
5 {\Gamma}^{hh, \,p_3}_{11,11} + {\Gamma}^{hh, \,p_3}_{00,00} + 10 {\Gamma}^{hh, \,p_3}_{01,11} +
5 {\Gamma}^{hh, \,p_3}_{00,11}  \nonumber \\
&& + {\Gamma}^{hh, \,p_3}_{11,00} +2 {\Gamma}^{hh, \,p_3}_{01,00} - 2 {\Gamma}^{hh, \,p_3}_{11,01} -
2 {\Gamma}^{hh, \,p_3}_{11,01} \nonumber \\
&& - 2 {\Gamma}^{hh, \,p_3}_{00,01} - 4 {\Gamma}^{hh, \,p_3}_{01,01}\, , \nonumber
\end{eqnarray}
where the $(\v{k},k_F,\theta)$--dependence in all the functions are omitted for simplicity.
For the $\vec{\Lambda} pp  \rightarrow  npp$ decay we obtain instead:
\begin{eqnarray}
\label{gamma2pp}
\Gamma^{hh, \,p_1, \, p_3}_{pp} & = & 4 (
{\Gamma}^{hh, \,p_1, \, p_3}_{11,11} + {\Gamma}^{hh, \,p_1, \, p_3}_{11,00} +
2 {\Gamma}^{hh, \,p_1, \, p_3}_{11,01})\, , \\
\Gamma^{hh, \,p_2, \, p_3}_{pp}  & = &
{\Gamma}^{hh, \,p_2, \, p_3}_{11,11} + {\Gamma}^{hh, \,p_2, \, p_3}_{00,00}
+ {\Gamma}^{hh, \,p_2, \, p_3}_{00,11} +
{\Gamma}^{hh, \,p_2, \, p_3}_{11,00} + 2 {\Gamma}^{hh, \,p_2, \, p_3}_{01,11} \nonumber \\
&& -2 {\Gamma}^{hh, \,p_2, \, p_3}_{01,00} + 2 {\Gamma}^{hh, \,p_2, \, p_3}_{11,01} +
2 {\Gamma}^{hh, \,p_2, \, p_3}_{00,01} - 4 {\Gamma}^{hh, \,p_2, \, p_3}_{01,01}\, . \nonumber
\end{eqnarray}

The last step is to integrate
over $(\v{k},k_F)$ in order to implement the local density approximation as
seen in Eq.~(\ref{decwpar3}).  We have, then:
\begin{eqnarray}
\label{g2hh}
\Gamma^{hh}_{np}(\theta) & \equiv & \Gamma^{hh, \,p_1}_{np}(\theta) +
\Gamma^{hh, \,p_2}_{np}(\theta) +\Gamma^{hh, \,p_3}_{np}(\theta)\, , \\
\Gamma^{hh}_{pp}(\theta) & \equiv & \Gamma^{hh, \,p_1, \, p_3}_{pp}(\theta) +
\Gamma^{hh, \,p_2, \, p_3}_{pp}(\theta)\, . \nonumber
\end{eqnarray}

In the Appendix, we show the derivation of some of the other contributions.
Once one normalizes per non--mesonic weak decay, these expressions are
inserted in Eq.~(\ref{np1f}) and (\ref{np1f-2}) to obtain the final result
for $N^{2N+FSI}_{p}(\theta)$ [see Eq.~(\ref{np1f-0})].

Before presenting the results, we anticipate some elements which emerge from the
obtained analytical expressions and the numerical
calculation. First, the $ph$ contribution of Fig.~\ref{2Nfsi} turns out to be negligibly small.
In addition, we have checked that the behavior of $N^{2N+FSI}_{p}(\theta)$ is
approximately linear in $\cos\theta$. We expect this result because for
the dominant $hh$ and $pp$ terms [see Eq.~(\ref{spphh})] the spin dependence
which generates the asymmetry is given by the same function
${\cal S}^{dir, \, \uparrow}_{\tau \tau'}(q)$
of Eq.~(\ref{tdir}) which enters the calculation of the intrinsic asymmetry.
This allows us to obtain the final expression for the asymmetry as follows:
\begin{equation}
\label{asimt_m}
a^{1N+2N+FSI}_{\Lambda} = \frac{N^{1N+2N+FSI}_{p}(0^{0}) - N^{1N+2N+FSI}_{p}(180^{0}) }
{N^{1N+2N+FSI}_{p}(0^{0}) + N^{1N+2N+FSI}_{p}(180^{0})}\, .
\end{equation}
Our predictions for $a^{1N+2N+FSI}_{\Lambda}$ can be directly compared with the data obtained for
the observable asymmetry $a^{\rm M}_\Lambda$.

\section{Numerical Results}
\label{numerics}
The weak transition potential $V^{\Lambda N\to NN}$ of Eq.~(\ref{intlnnn})
is described in terms of the usual one--meson--exchange
(OME), together with the uncorrelated and correlated
two--pion--exchange, which was shown to have a very important effect on the
asymmetry \cite{Ch07}.
The OME potential is represented by the exchange of $\pi$,
$\eta$, $K$, $\rho$, $\omega$ and $K^*$ mesons within the formulation
of~\cite{pa97}, with values of the coupling constants and cutoff parameters
taken from \cite{na77} (Nijmegen89) and \cite{st99} (Nijmegen97f).
We present results for both Nijmegen89 and Nijmegen97f weak transition
potentials for the
following reason. The adopted two--pion--exchange potential was introduced
within a chiral unitary approach in~\cite{ji01}, together with an important
compensatory $\omega$-exchange contribution with a
 a $\Lambda N \omega$ parity--conserving coupling,
$g^{\omega}_{\Lambda N \omega}=3.69 \, G_{F} m^{2}_{\pi}$, which is the same of the
Nijmegen89 potential. At variance, in the Nijmegen97f potential one has
$g^{\omega}_{\Lambda N \omega}=0.17 \, G_{F} m^{2}_{\pi}$. One thus expects
a difference between the Nijmegen89 and Nijmegen97f results for the asymmetry.
Although the Nijmegen89-based weak potential was the one originally employed in
conjuction with the two--pion--exchange mechanism, in a set of recent
contributions we have used the Nijmegen97f potential and we believe that it is
of interest to
discuss this particular parametrization too.

For the nucleon--nucleon strong interaction $V^{N N}$ of Eq.~(\ref{intnn1}) we have used the
Bonn potential~\cite{ma87} in the framework of the parametrization
of~\cite{br96}, which contains the exchange of $\pi$,
$\rho$, $\sigma$ and $\omega$ mesons and neglects the $\eta$ and
$\delta$ mesons. We present results for $^{12}_{\Lambda}$C, where the
hyperon is assumed to decay from the $1s_{1/2}$ orbit of a harmonic
oscillator well with frequency $\hbar \omega = 10.8$ MeV adjusted to the experimental
energy separation between the $s$ and $p$ $\Lambda$--levels in
$^{12}_{\Lambda}$C \cite{exp_hashimoto}.

\subsection{Non--mesonic decay rates}
\label{numerics-rates}

The two--pion--exchange potential is introduced in our microscopic approach
for the first time here. It is thus important to start our discussion showing
the numerical results for the non--mesonic
weak decay widths.
These rates are given in Table~\ref{gammas} for the two transition potentials,
Nijmegen89 and Nijmegen97f, without (OME) and with (OME$+2\pi$)
the two--pion--exchange contribution.
Let us start by discussing the independent
rates $\Gamma_n$, $\Gamma_p$ and $\Gamma_2$. For $\Gamma_n$ and $\Gamma_2$ all predictions
agree with data within error bars; instead, apart from the OME result with the
Nijmegen97f potential, our predictions overestimate the data for $\Gamma_p$.
The origin of the agreement for $\Gamma_n$ and the disagreement for $\Gamma_p$ is not
known. However, it is the same which leads to a good description of experimental emission
spectra involving only neutrons and an overestimation of data on spectra involving at least one
proton. This is proved by the comparison of all the theoretical approaches
\cite{ga03,Ba10,ba11} to the single and double--coincidence nucleon emission distributions
with the corresponding KEK \cite{Kim09} and FINUDA \cite{FINUDA} data.
These nucleon spectra are the real observables in non--mesonic
decay, while the experimental values of the partial decay rates $\Gamma_n$, $\Gamma_p$, etc,
are obtained after a deconvolution of the FSI effects contained in the measured spectra.
The disagreement on the spectra is thus the fundamental problem, which also affects
the above disagreement on the $\Gamma_p$ rate.
\begin{table}[h]
\begin{center}
\caption{The non--mesonic decay widths predicted for $^{12}_\Lambda$C
(in units of the free decay rate). The most recent data, from KEK--E508~\protect\cite{Kim09}
and FINUDA~\protect\cite{FINUDA,FINUDA2}, are also given.}
\label{gammas}
\begin{tabular}{lccccccccc}   \hline\hline
 &   \mc {2}{c}{Nijmegen89} && \mc {2}{c}{Nijmegen97f}  &\\ \cline{2-3} \cline{5-6}
 & ~~OME~~ & OME$+2 \pi$ && ~~ OME ~~& OME$+2 \pi$  & KEK--E508 \protect\cite{Kim09} &
FINUDA \protect\cite{FINUDA} & FINUDA \protect\cite{FINUDA2} \\ \hline
$\Gamma_n$                 & $0.19$ & $0.15$  &&  $0.16$  & $0.23$ &  $0.23\pm 0.08$  &  &  \\
$\Gamma_p$                 & $0.65$ & $0.61$  &&  $0.47$  & $0.65$ &  $0.45\pm 0.10$  &  &  \\
$\Gamma_1$                 & $0.84$ & $0.76$  &&  $0.63$  & $0.88$ &  $0.68\pm 0.13$  &  &  \\
$\Gamma_2$                 & $0.17$ & $0.26$  &&  $0.17$  & $0.37$ &  $0.27\pm 0.13$  &  &  \\
$\Gamma_{\rm NM}$          & $1.01$ & $1.02$  &&  $0.80$  & $1.25$ &  $0.95\pm 0.04$  &  &  \\
$\Gamma_n/\Gamma_p$        & $0.29$ & $0.25$  &&  $0.34$  & $0.35$ &  $0.51\pm 0.13\pm 0.05$  &  &  \\
$\Gamma_2/\Gamma_{\rm NM}$ & $0.17$ & $0.26$  &&  $0.21$  & $0.30$ &  $0.29\pm 0.13$  &  $0.24\pm 0.10$ &
$0.21\pm 0.07^{+0.03}_{-0.02}$ \\
\hline\hline
\end{tabular}
\end{center}
\end{table}

From Table~\ref{gammas} we also see that the effect of the two--pion--exchange
potential is different when added to the Nijmegen89 and Nijmegen97f OME potentials.
This is due to the different values of the $g^{\omega}_{\Lambda N \omega}$ coupling
constant previously discussed. While there is a moderate reduction of
$\Gamma_n$ and $\Gamma_p$ for Nijmegen89, an increase of these decay rates is
observed in the case of Nijmegen97f. The behaviour in this later case agrees
with what was found in~\cite{Ch07,chumi12} with the same weak transition
potential \footnotemark{\footnotetext{We note that the results of~\cite{Ch07} have been
recently revised using more realistic form factors \cite{chumi12}. Although
the numerical values have slightly changed, the qualitative aspects of the
two--pion--exchange mechanism remain the same.}}. The addition of the
two--pion--exchange
potential increases substantially the value of $\Gamma_{2}$ for both the
Nijmegen89 and Nijmegen97f, the effect being stronger for the later case.
There is a certain dispersion
among the results obtained with the different potentials. However, considering the
big error bars for data and the mentioned discrepancy on
the proton emission spectra, we believe that all the four potential models
of Table~\ref{gammas} should be also considered in the analysis of the asymmetry
parameter.

\subsection{The asymmetry parameter}
\label{numerics-asy}

We start by discussing the intrinsic asymmetry $a^{1N}_\Lambda$.
In Table~\ref{asim_results} we compare our predictions (first four lines) with
the results reported in the literature (last four lines), where the updated
results of the finite nucleus calculation of~\cite{Ch07} have been listed.

We obtain a rather sizable asymmetry
parameter for the OME Nijmegen89 and Nijmegen97f models, in agreement with
other works, especially with the nuclear
matter result of~\cite{du96}. Note that our OME results are more moderate
than any of the values found by calculations performed in finite nuclei. This
is probably due to the more extended Fermi motion effects in nuclear matter.
The inclusion of the two--pion--exchange mechanism
strongly decreases the absolute value of $a^{1N}_\Lambda$, especially for the
Nijmegen97f model, in agreement with what was found by the finite nucleus
calculation of Chumillas {\it et al.}\cite{Ch07}.

Finally, we note that
the appreciable difference for the intrinsic asymmetry results evaluated within
our two ${\rm OME}+2\pi$ models is a consequence of the different
values for $g^{\omega}_{\Lambda N \omega}$ of the Nijmegen89 and Nijmegen97f OME
potentials.

\begin{table}[h]
\begin{center}
\caption{Theoretical determinations of the intrinsic asymmetry parameter
for $^{12}_{\Lambda}$C. The calculations reported from the literature were performed
within shell model approaches, except for the nuclear matter result of
Dubach {\it et al.}. The resuls of the model of Chumillas {\it et al.}
\cite{Ch07}, correspond to the ones updated in \cite{chumi12}.}
\label{asim_results}
\begin{tabular}{lc}   \hline\hline
Model & $a^{1N}_{\Lambda}$($^{12}_{\Lambda}$C) \\ \hline
OME (Nijmegen89)& $-0.39$ \nonumber \\
OME (Nijmegen89) $+\, 2\pi$ & $-0.23$ \nonumber \\
OME (Nijmegen97f)& $-0.35$ \nonumber \\
OME (Nijmegen97f) $+\, 2\pi$ & $-0.071$ \nonumber \\ \hline
Ref. and Model &  \\ \hline
Dubach {\it et al.} \cite{du96},
OME (NM)& $-0.44$ \nonumber \\
Parre\~no and Ramos \cite{pa02},
OME & $-0.55$ to $-0.73$\nonumber \\
Barbero {\it et al.} \cite{bar05},
OME& $-0.53$ \nonumber \\
Chumillas {\it et al.} \cite{Ch07,chumi12}, OME & $-0.48$ \nonumber
\\
Chumillas {\it et al.} \cite{Ch07,chumi12}, OME$+2\pi$ & $-0.0062$ \nonumber\\
\hline\hline
\end{tabular}
\end{center}
\end{table}

Before moving into the new effects explored in this work, let us comment on the
fact that our microscopic calculation of the asymmetry parameter takes care of
the two isospin
channels depicted in Fig.~\ref{exccon}. In fact, these contributions are
automatically encoded within the antisymmetric character of the final
two-nucleon wave-function used in finite-nucleus calculations of the
weak decay. However, the diagrammatic approach employed here is useful
in the sense that it allows one to keep track of the importance of the different
contributions to the asymmetry parameter. It turns out that diagram $(a)$ is the
dominant contribution. This can be explained as follows.
Let us denote with $\v{p}_{a}$ ($\v{p}_{b}$)
the momentum carried by the proton in diagram $(a)$ ($(b)$).
From the kinematics of diagrams $(a)$ and $(b)$ we have:
\begin{eqnarray}
\label{kinematics}
\v{p}_{a} & = &  \v{k}-\v{q} \cong  -\v{q}\, ,\\
\v{p}_{b} & = &  \v{h}+\v{q} \cong  \v{q}\, ,\nonumber
\end{eqnarray}
where, for the purpose of this explanation, it is a good approximation
to assume that $\v{k}$ and $\v{h}$ are much smaller than $\v{q}$. The
different sign but similar magnitude of $\v{p}_{a}$ and $\v{p}_{b}$ means that
a negative asymmetry from the charge-exchange diagram $(a)$
is reduced in magnitude by a positive asymmetry from diagram $(b)$. The
competition between the diagrams $(a)$ and $(b)$ produces a
reduction in the absolute value of the asymmetry. This type of analysis will be
particularly useful for the $2N$ and FSI effects discussed
below.

In Table~\ref{asim_results4} we present our predictions for the asymmetry
when $2N$ and FSI--induced decays are considered together
with $1N$ decays. Since any experiment is affected by a kinetic energy
threshold for proton detection, $E_{th}$, results are also given
for different values of $E_{th}$.

\begin{table}
\begin{center}
\caption{Effect of the nucleon--nucleon strong interaction on the asymmetry parameter for
$^{12}_{\Lambda}$C.
In addition to the contribution from the $\vec \Lambda p\to np$ weak decay leading
to the intrinsic asymmetry, we consider the action of $2N$ and FSI--induced decays.
The most recent data are also shown.}	
\label{asim_results4}
\begin{tabular}{llccccc}   \hline\hline
 &  & \mc {2}{c}{Nijmegen89} && \mc {2}{c}{Nijmegen97f}  \\ \cline{3-4} \cline{6-7}
~$E_{th}$~(MeV)~&~~Asymmetry~~&
 ~~~~OME~~~~ & ~OME$+2 \pi$ ~&& ~~~~ OME ~~~~& ~OME$+2 \pi$ ~\\ \hline
 ~~~0   &  $a^{{1N}}_{\Lambda}$    & $-0.386$  & $-0.225$   && $-0.352$ &
$-0.071$ \\ \hline
~~~0 &$a_{\Lambda}^{{1N} +2N}$         & $-0.366$  &   $-0.212$ && $-0.318$ & $ -0.063$ \\
&$a_{\Lambda}^{{1N} +FSI}$         & $-0.184$  &   $-0.009$ && $-0.043$ & $~~~0.082$ \\
&$a_{\Lambda}^{{1N} +2N+FSI}$     & $-0.234$  &   $-0.071$ && $-0.132$ & $~~~0.032$\\ \hline
~~30&  $a_{\Lambda}^{{1N} +2N}$  & $-0.355$  &   $-0.197$ && $-0.307$ & $ -0.060$ \\
&$a_{\Lambda}^{{1N} +FSI}$         & $-0.149$  &   $-0.003$ && $-0.034$ & $~~~0.096$ \\
&$a_{\Lambda}^{{1N} +2N+FSI}$     & $-0.196$  &   $-0.056$ && $-0.115$ & $~~~0.037$\\ \hline
~~50&  $a_{\Lambda}^{{1N} +2N}$  & $-0.319$  &   $-0.149$ && $-0.255$ & $ -0.049$ \\
&$a_{\Lambda}^{{1N} +FSI}$         & $-0.123$  &   $~~~0.019$ && $~~~0.014$ & $ ~~~0.112$ \\
&$a_{\Lambda}^{{1N} +2N+FSI}$     & $-0.156$  &   $-0.018$ && $-0.058$ & $~~~0.069$\\ \hline
~~KEK--E508\cite{mar06} &        & &&&  \mc {2}{c}{$-0.16\pm0.28^{+0.18}_{-0.00}$}\\ \hline \hline
\end{tabular}
\end{center}
\end{table}

The values obtained for the asymmetry depend on two important
effects. The first one is
the dynamics of the weak transition. In particular, one can consider or not the
two--pion--exchange potential. This has been analyzed in detail in~\cite{al05,Ch07}, and our
results confirm those findings. Moreover,
the asymmetry depends on what we call
``kinematic effect''. The introduction of the different $2N$ and FSI contributions
enlarges the available phase, leading to a particular kinematics for each contribution;
the weight imposed by the nucleon--nucleon strong interaction on the different kinematics
(and also the restrictions due to $E_{th}$) modifies the relation between
$N_{p}(0^{0})$ and $N_{p}(180^{0})$. The microscopic model is
particularly suitable for the study of this kinematic effect.
Note that the division between the dynamic and the kinematic effects is possible because
the spin summation representing the interference between parity--violating and
parity--conserving terms of the transition potential has the same expression,
given by Eq.~(\ref{tdir}), for the intrinsic asymmetry and for the dominant $2N$ and FSI
contributions to the observable asymmetry.

\begin{figure}[h]
\centerline{\includegraphics[scale=0.5]{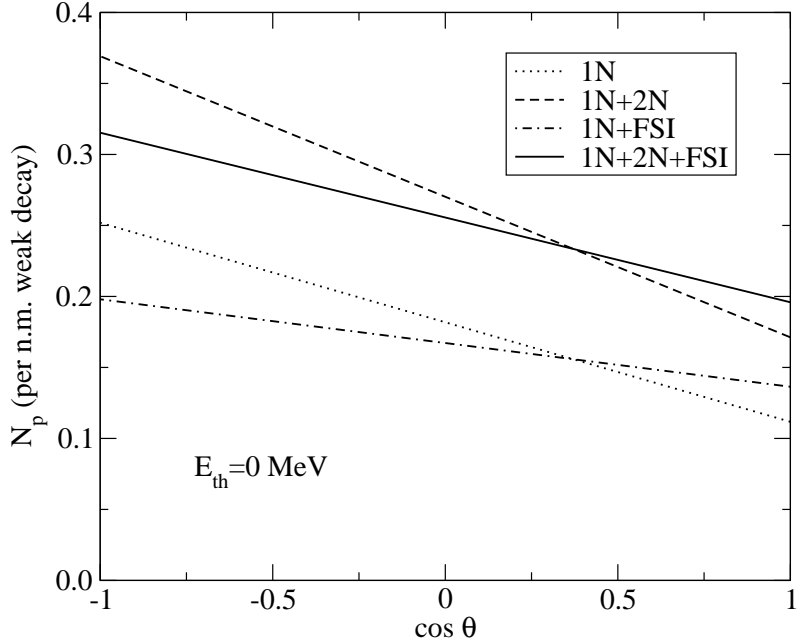}}
\caption{$N_p^{1N}$ (dotted line), $N_p^{1N+2N}$ (dashed
line), $N_p^{1N+FSI}$ (dash-dotted line) and total
$N_p^{1N+2N+FSI}$ (solid line), as functions of $\cos\theta$, in the
the case of the OME Nijmegen89 weak transition potential.}
\label{fig:asym1}
\end{figure}

It is instructive to recall that the value of the asymmetry parameter is a consequence
of a delicate balance between parity--conserving and parity--violating amplitudes
governed by the dynamics of the weak decay mechanism, and also depends on the
phase space allowed for the emitted nucleons which might be enhanced or
decreased in some places by strong interaction effects or kinematical cuts.
Any new contribution to the decay process will introduce
changes in the number of protons emitted parallel, $N_p(0^0)$,
and antiparallel, $N_p(180^{0})$, to the polarization
axis, therefore affecting the value of the asymmetry which is determined
by the difference $N_p(0^0)-N_p(180^{0})$
measured relative to the sum $N_p(0^0)+N_p(180^{0})$, as seen in
Eq.~(\ref{asimt_m}). It is therefore illustrative to represent the
function $N_p(\theta)$ as a function of $\cos\theta$, as seen in
Fig.~\ref{fig:asym1} for the OME Nijmegen89 model, including the $1N$ induced
decays (dotted line), adding the $2N$-induced modes (dashed line), adding only
FSI effects (dash-dotted line), and finally incorporating all the contributions
together (solid line). Similar plots are obtained for the other three potential
models employed in this work.

It is clear that the $2N$ mechanism enhances the number of emitted protons at
all emitted angles but having a slight preference for directions opposite
to the polarization axis, hence the size of the
slope of the dashed line in Fig.~\ref{fig:asym1} is a little bit larger than
that of the dotted line. This would increase the magnitude of the asymmetry but
the larger number of protons gives finally rise to a
slight decrease, as seen in Table~\ref{asim_results4}. Diagrammatically
speaking, we know that, for $1N$ decays, the asymmetry
receives the main (negative) contribution when the proton is
attached to the $\Lambda$
vertex (diagram $(a)$ of Fig.~\ref{exccon}). The other (positive) contribution,
with a neutron
outgoing from the $\Lambda$ vertex (diagram $(b)$ of Fig.~\ref{exccon}), tends
to reduce
the absolute value of the asymmetry.
In the case of $2N$ decay diagrams, one has $npp$ and $nnp$ final states,
where the
proton(s) can be located at the $\Lambda$ vertex or in any of the two remaining
positions.
It is the increased number of positions for the final proton(s) that produces a
further reduction
(although small) of the asymmetry parameter.
We note that the small effect of $2N$ decays on the asymmetry parameter
corroborates the
assumption done in~\cite{al05}.

As far as FSI effects are concerned, we observe in Fig.~\ref{fig:asym1}
that they remove antiparallel protons and, on the other hand, more strength is
added at parallel kinematics. Since the total number of protons is almost
unchanged, this reduction of slope observed for the dot-dashed line
also translates in a subtantial decrease in the magnitude of the asymmetry.
In order to analyze further the different FSI contributions, it
is convenient to write:
\begin{equation}
\label{fsi_split}
N^{FSI}_{p}(\theta) \equiv N^{2p1h}_{p}(\theta) + N^{3p2h}_{p}(\theta)\, ,
\end{equation}
where each of the two terms on the rhs receive contributions from each of the
diagrams in Fig.~\ref{2Nfsi}, by cutting on $2p1h$ or $3p2h$ states,
respectively. By construction, $N^{2p1h}_{p}(\theta)$
originates from a QIT between $1N$ and FSI--induced decays, while
$N^{3p2h}_{p}(\theta)$ may come either from a $2N$--FSI QIT term or
from a pure FSI--induced decay. The microscopic model allows us to inspect the
behavior of each term. We find that the term $N^{3p2h}_{p}(\theta)$ is
positive--definite and has a similar
behavior to the one already discussed for $N^{2N}_{p}(\theta)$, i.e.
slightly more protons are emitted antiparallel to the polarization axis.
On the contrary,
$N^{2p1h}_{p}(\theta)$ turns out to be negative while its
kinematic behavior is very similar to the $1N$-induced decays, which
produce a large negative asymmetry parameter. Therefore, the effect of the
negative $N^{2p1h}_{p}(\theta)$ contributions goes in the direction of inverting
this behavior, giving rise to a subtantial decrease in the size of the asymmetry
or even reverting its sign, as in the case of
the Nijmegen97f$+2\pi$ model.

\begin{figure}[h]
\centerline{\includegraphics[scale=0.5]{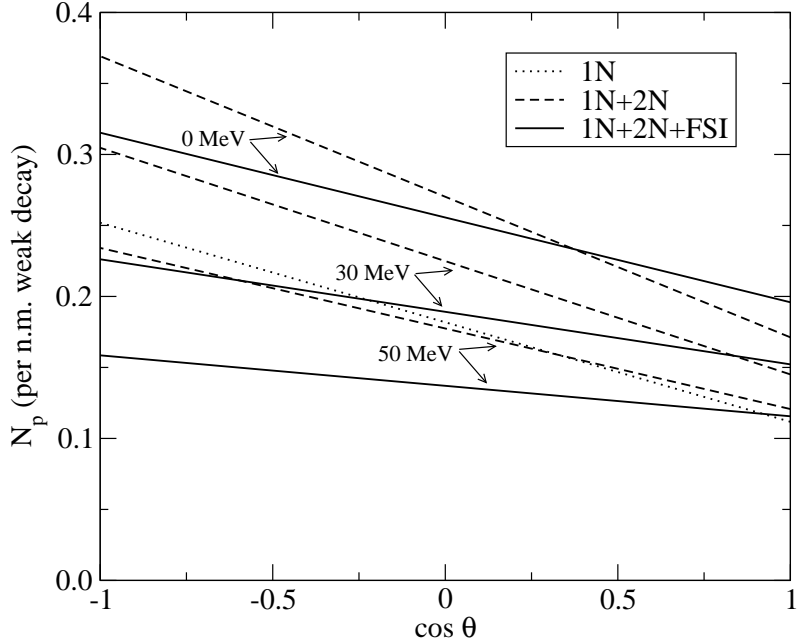}}
\caption{The functions $N_p^{1N+2N}$
(dashed lines) and
$N_p^{1N+2N+FSI}$ (solid lines) for different energy cuts, $E_{th}=0$,30
and 50 MeV, in the
the case of the OME Nijmegen89 weak transition potential. The dotted line
corresponds to $N_p^{1N}$ at $E_{th}=0$ MeV.}
\label{fig:asym2}
\end{figure}

We now pay attention to the behavior of the asymmetries of
Table~\ref{asim_results4} with the enegy cut $E_{th}$. We observe that the size
of the asymmetry
$a_{\Lambda}^{1N+2N}$  decreases slightly for increasing
$E_{th}$. This reduction can be explained from a microscopic point of view by
inspecting the momentum
distribution predicted by our approach for the three particles,
$p_1$ (particle outgoing from the $\Lambda$ vertex), $p_2$ and $p_3$, stemming
from $2N$
decays, shown in Fig.~12 of \cite{Ba10}. Particles are named in that figure with
the same notation as in Fig.~\ref{2Nd} of the present work.
The distributions for particles $p_1$ and $p_3$ are very similar to each
other
and are peaked at a lower momentum than the distribution for $p_2$.
Due to isospin reasons, the main (negative) contribution to the asymmetry
is obtained when a proton is located in $p_{1}$,
while protons in $p_{2}$ and/or $p_{3}$ reduce the magnitude of the asymmetry.
The effect
of $E_{th}$ is to reduce the importance of the particle $p_{1}$ with respect
to $p_{2}$. This explains the reduction in the magnitude of $a_{\Lambda}^{{1N} +
2N}$ for increasing $E_{th}$.
Note, however, that the decrease is much stronger in the case of the
$a_{\Lambda}^{1N+FSI}$ asymmetry, and this is also the behavior of the
complete calculation, $a_{\Lambda}^{1N+2N+FSI}$. In order to understand this
behavior, we recall that the energy cut removes nucleons, making
$N_p(0^0)+N_p(180^0)$ smaller and, consequently, the magnitude of the asymmetry
larger. But the final consequence of the $E_{th}$ cut on
the asymmetry will be determined by whether the increase in size due
to the removal of nucleons is counterbalanced by the changes in the
slope $N_p(0^0)-N_p(180^0)$. In Fig.~\ref{fig:asym2}
we show the effect of this cut for $N_p^{1N+2N}(\theta)$ (dashed lines)  and
$N_p^{1N+2N+FSI}(\theta)$ (solid lines) as functions of $\cos\theta$, for the
OME Nijmegen89 model. We clearly see a reduction in the number of protons as
well as a decrease in the slope with increasing $E_{th}$ . These two
effects modify the asymmetry parameter in opposite ways and the results of
Table~\ref{asim_results4} show that,  within our models,
the reduction of the
asymmetry due to the decrease in the slope dominates over the increase
associated to the removal of particles. The reduction in the slope
is much more pronounced for the FSI contributions. The final result is that
we observe a substantial reduction in the magnitude of the
asymmetry $a_{\Lambda}^{1N+2N+FSI}$ with an increasing energy cut.

This behavior contrasts with the INC results of~\cite{al05,Ch07},
where, for increasing $E_{th}$, the size of the asymmetry $a^{1N+FSI}_\Lambda$
increases
and tends to the intrinsic value. We note that the INC model for FSI originates
from
a semi--classical description which has an intuitive interpretation. Nucleons
are
tracked in their way out of the nucleus as classical particles.
Sometimes a nucleon leaves the nucleus without any interaction with the medium,
in other
cases it scatters one or more times with the other bound nucleons. Therefore, a
nucleon
emerging from an elementary non--mesonic decay can change momentum, direction
and charge,
other nucleons can be emitted as well, etc.
Clearly, the random character of these FSI processes is responsible for the
strong reduction of $a^{1N+FSI}_\Lambda$  by about a factor two or more
with respect to the intrinsic asymmetry
$a^{1N}_\Lambda$ \cite{al05,Ch07}.
The introduction of an energy cut $E_{th}$ affects
mainly those nucleons which have suffered scattering processes.
For increasing $E_{th}$, the nucleons coming from elementary
decays (not affected by FSI) become dominant and the asymmetry tends to the
intrinsic value. This is also reflected in Fig. 2 of~\cite{al05} by the
tendency of $N_p^{1N+FSI}(\theta)\mid_{E_{th}}$ to move towards
$N_p^{1N}(\theta)$ as the energy cut is increased, both functions becoming very
similar (in size and slope) at around
$E_{th}=50$ MeV.

The situation for our microscopic approach is different as it is based on
quantum mechanics, where QIT play an important role.
It was shown in~\cite{ba11} that the $2p1h$ and $3p2h$ terms of
the proton kinetic energy
spectra $N_p^{FSI}(T_{p})=N_p^{2p1h}(T_{p})+N_p^{3p2h}(T_{p})$ have a different
behavior
from each other. While $N_p^{3p2h}(T_{p})$ gives a positive distribution,
has its maximum for $T_p=0$ and decreases for increasing $T_p$,
the QIT $N^{2p1h}(T_{p})$ is a negative bell--shaped
distribution with
the minimum at $T_{p} \cong 80$~MeV. Thus, a non--vanishing energy cut $E_{th}$
appreciably reduces $N_p^{3p2h}$ while leaving
$N_p^{2p1h}$ almost unchanged, and this later contribution is the one
producing a significant decrease in the slope of $N_p^{1N+2N+FSI}(\theta)$.
Therefore, as  $E_{th}$ is increased, the magnitude of the asymmetry parameter
decreases.

We end our discussion by comparing our results with experiment.
The asymmetry data reported in Table~\ref{asim_results4} was obtained by
KEK--E508 for a kinetic energy threshold $E_{th}$ of about 30 MeV.
Despite the noticeable differences among the whole set of predictions,
they are all compatible with experiment due to the large error bar of data.
By considering our results for  $E_{th}=30$ MeV and the central value of the experimental data,
the best agreement is obtained for both Nijmegen89 and Nijmegen97f models
when the two--pion--exchange potential is not included.
Certainly, our calculation shows that the effect of the two--pion--exchange
is very important in asymmetry calculations, but to establish definite conclusions on the effect
of this potential more detailed studies are required. In general, the addition
of any new
contribution to the weak transition potential has to be done consistently with
the rest of the potential itself (which might require some readjustment to
reproduce the observables) and with the approach adopted in the calculation.
However, in order to obtain fruitful information from these studies new and
more precise data is needed to constrain the unknown parameters of the weak
decay models.

\section{Remarks and Conclusions}
\label{conclusions}

We have discussed a microscopic diagrammatic formalism to evaluate the
asymmetry in the distribution of protons emitted in the non--mesonic decay of
polarized hypernuclei.
The calculation is performed in nuclear matter and then extended to finite hypernuclei
($^{12}_\Lambda$C) by means of the local density approximation.
Our approach takes into account both the $2N$ decay mechanism and the nucleon FSI
in a unified many--body scheme. The effect of the $2N$ decays on the
asymmetry parameter is evaluated here for the first time.
The present work is also the first one to implement the FSI on the asymmetry
parameter by means of a quantum-mechanical microscopic approach.
In addition to the usual OME weak transition
potentials, which we take from the Nijmegen89 and Nijmegen97f parametrizations,
we have also considered the effect of the two--pion--exchange potential
introduced in~\cite{ji01}. We give results for both the intrinsic asymmetry
parameter, $a^{1N}_\Lambda$, and for the asymmetry parameter modified by the
$2N$-induced mechanisms and FSI effects, $a^{1N+2N+FSI}_\Lambda$,
which is the one that can be compared to the observed asymmetry $a^{\rm
M}_\Lambda$.

While the effect of $2N$ is predicted to be rather limited,
the nucleon FSI turned out to be very important: they reduce the
magnitude of the asymmetry parameter, making all
the weak transition potential models adopted in this work capable of
describing consistently the experimental data for $a^{\rm M}_\Lambda$ and for
the non--mesonic weak decay rates.
In particular,
the large error bars in the observable asymmetry do not allow us to determine which of the
two potential models, OME or OME$+2\pi$, provides the best description of the experiments.

To the best of our knowledge, the only former work which evaluated the intrinsic asymmetry in
nuclear matter is due to Dubach et al.~\cite{du96}, where an approximate scheme rather
different from ours (neglecting $2N$ decays and nucleon FSI) was employed. The action of FSI
was considered within a semi--classical description in~\cite{al05,Ch07}, by means of an INC
model. In a former calculation \cite{ba11}, it was shown that the
INC model and the present microscopic approach provide similar results for the
nucleon emission spectra in the
non--mesonic weak decay  of unpolarized $\Lambda$ hypernuclei.  Our results for
$a^{1N}_\Lambda$ and $a^{1N+FSI}_\Lambda$ with a vanishing proton kinetic
energy cut, $E_{th}=0$, fairly agree with each other too. However, the situation
changes for non--vanishing values of $E_{th}$. For increasing $E_{th}$, the
negative asymmetry of the INC model increases in magnitude,
while a decrease is observed in the microscopic model.
One should note that
the two schemes represent rather different approaches to the problem of dealing
with nuclear
correlations after the weak decay takes place. The microscopic model of the
present work provides a reliable method that can be improved systematically. It
however ignores multinucleon processes that are accounted for, semiclassically
and via multi-step processes, in the INC model. To determine which is the most
realistic approach, an accurate experimental determination of the asymmetry
parameter, possibly
exploring its $E_{th}$--dependence, would certainly be welcome.

One should always keep in mind that the main motivation in the study of the non--mesonic
weak decay of hypernuclei is to extract information on strangeness--changing baryon--baryon
interactions. The understanding of the $\Gamma_{n}/\Gamma_{p}$ ratio and the asymmetry
parameter suggests that a fairly reasonable knowledge of non--mesonic decay has
been achieved. However, we have obtained agreement with all the experimental
data employing different parametrizations of the weak transition potential. Due
to the lack of precise data for the asymmetry parameter, we find that the role
of the two--pion--exchange mechanism, which was essential to reproduce this
observable in some models \cite{Ch07}, can not even be firmly established here.
In any case, what is certain is the agreement with the set of data can only be
achieved after a proper development of
approaches that take care of nucleon FSI. Due to the special nature of
the in--medium non--mesonic weak decay, these are complex models, but
they are required to establish a link between theory and experiment.

Finally, we recall that there still remains
an important disagreement between theory an experiment for the hypernuclear non--mesonic
weak decay: theoretical evaluations of nucleon emission spectra involving protons
strongly overestimate the experimental distributions.
This discrepancy may not be isolated but hidden behind the errors bars in
the data for the decay rates. An additional aspect that has not yet been studied but
which could lead to a non--negligible contribution to the nucleon spectra is the inclusion
of the $\Delta(1232)$--resonance in our many--body Feynman diagram scheme.
We intend to study this problem in the future.

\section*{Acknowledgments}
This work has been partially supported by the CONICET and ANPCyT, Argentina,
under contracts PIP 0032 and PICT-2010-2688, respectively,
by the contract FIS2008-01661 from MICINN
(Spain) and by the Ge\-ne\-ra\-li\-tat de Catalunya contract 2009SGR-1289. We
acknowledge the support of the European Community-Research Infrastructure
Integrating Activity ``Study of Strongly Interacting Matter'' (HadronPhysics2,
Grant Agreement n. 227431) under the Seventh Framework Programme of EU.

\section*{Appendix}
\label{APPENDA}
Here we present the explicit expressions needed in the evaluation of
$N_p^{2N}(\theta)$ starting from the Feynman diagrams $pp$ and $ph$ in Fig.~\ref{2Nfsi}.
We omit the derivation of the expressions for $N_p^{FSI}(\theta)$ obtained from
the same diagrams, as they can be obtained from the $N_p^{2N}(\theta)$ ones
after some simple changes: the spin--isospin structures are the same, as
well as the general expressions, except for some step functions and
energy denominators.

We begin with the contribution of the diagram $pp$ of Fig.~\ref{2Nfsi}. First, we
introduce the partial, isospin--dependent decay widths for $\vec \Lambda np\to nnp$:
\begin{eqnarray}
\label{gamppp1}
\Gamma^{pp, \, p_1}_{\tau \tau'; \tau_{N} {\tau'}_{N}}(\v{k},k_F,\theta) & =
& \frac{(G_F m_{\pi}^2)^{2}}{(2 \pi)^{5}}
\left(\frac{f_{\pi}^2}{4 \pi}\right)^{2}  \frac{1}{m_{\pi}^4}
\frac{2}{(2 \pi)^2} \;
\int d  \v{q}
\int d  \v{t} \int d \v{h} \int d \v{h'} \,
\theta(q_{0}) \\
&& \times \theta(|\v{k}-\v{q}| - k_F)
\theta(k_F - |\v{h}+\v{q}|) \, \theta(|\v{h}-\v{t}+\v{q}| - k_F)
\theta(k_F - |\v{h}|)
\nonumber \\
&&
\times \theta(|\v{h'}+\v{t}| - k_F) \,
\theta(k_F - |\v{h'}|) \; \delta(\cos\theta-(\v{k-q})_{z}/|\v{k-q}|)
\nonumber \\
&&
\times \delta(q_0 - (E_N(\v{h}-\v{t}+\v{q}) - E_N( \v{h})+E_N(\v{h'}+\v{t})-E_N(\v{h'}))
\nonumber \\
& &
\times \frac{{\cal S}^{pp}_{\tau \tau'; \tau_{N} {\tau'}_{N}}(q,t) }
{(E_N(\v{h}-\v{t}+\v{q})-E_N(\v{h}+\v{q})+E_N(\v{h'}+\v{t})-E_N(\v{h'}))^{2}}\, ,\nonumber \\
&& \nonumber \\
\Gamma^{pp, \, p_2}_{\tau \tau'; \tau_{N} {\tau'}_{N}}(\v{k},k_F,\theta) & =
& \Gamma^{pp, \, p_1}_{\tau \tau'; \tau_{N} {\tau'}_{N}}(\v{k},k_F,\theta)| \\
&&\delta(\cos\theta-(\v{k-q})_{z}/|\v{k-q}|) \rightarrow
\delta(\cos\theta-(\v{h-t+q})_{z}/|\v{h-t+q}|)\, , \nonumber \\
&& \nonumber \\
\Gamma^{pp, \, p_3}_{\tau \tau'; \tau_{N} {\tau'}_{N}}(\v{k},k_F,\theta) & =
& \Gamma^{pp, \, p_1}_{\tau \tau'; \tau_{N} {\tau'}_{N}}(\v{k},k_F,\theta)| \\
&&\delta(\cos\theta-(\v{k-q})_{z}/|\v{k-q}|) \rightarrow
\delta(\cos\theta-(\v{h'+t})_{z}/|\v{h'+t}|)\, , \nonumber
\end{eqnarray}
where $p_1$, $p_2$ and $p_3$ indicate the position of the final proton.
In a similar way, for the reaction $\vec{\Lambda} pp  \rightarrow  npp$ we have:
\begin{eqnarray}
\label{gamppp1p3}
\Gamma^{pp, \, p_1, \, p_2}_{\tau \tau'; \tau_{N} {\tau'}_{N}}(\v{k},k_F,\theta) & =
& \frac{(G_F m_{\pi}^2)^{2}}{(2 \pi)^{5}}
\left(\frac{f_{\pi}^2}{4 \pi}\right)^{2}  \frac{1}{m_{\pi}^4}
\frac{2}{(2 \pi)^2} \;
\int d  \v{q}
\int d  \v{t} \int d \v{h} \int d \v{h'} \,
\theta(q_{0}) \\
&& \times \theta(|\v{k}-\v{q}| - k_F)
\theta(k_F - |\v{h}+\v{q}|) \, \theta(|\v{h}-\v{t}+\v{q}| - k_F)
\theta(k_F - |\v{h}|)
\nonumber \\
&&
\times \theta(|\v{h'}+\v{t}| - k_F) \,
\theta(k_F - |\v{h'}|) \nonumber \\
&&\times
(\delta(\cos\theta-(\v{k-q})_{z}/|\v{k-q}|)+\delta(\cos\theta-(\v{h+q-t})_{z}
/|\v { h+q-t }|))/2
\nonumber \\
&&
\times \delta(q_0 - (E_N(\v{h}-\v{t}+\v{q}) - E_N( \v{h})+E_N(\v{h'}+\v{t})-E_N(\v{h'}))
\nonumber \\
& &
\times \frac{{\cal S}^{pp}_{\tau \tau'; \tau_{N} {\tau'}_{N}}(q,t) }
{(E_N(\v{h}+\v{q}-\v{t})-E_N(\v{h}+\v{q})+E_N(\v{h'}+\v{t})-E_N(\v{h'}))^{2}}\, , \nonumber\\
&&  \nonumber \\
\Gamma^{pp, \, p_1, \, p_3}_{\tau \tau'; \tau_{N} {\tau'}_{N}}(\v{k},k_F,\theta) & =
& \Gamma^{pp, \, p_1, \, p_2}_{\tau \tau'; \tau_{N} {\tau'}_{N}}(\v{k},k_F,\theta)| \\
&&\delta(\cos\theta-(\v{h+q-t})_{z}/|\v{h+q-t}|) \rightarrow
\delta(\cos\theta-(\v{h'+t})_{z}/|\v{h'+t}|), \nonumber \\
&& \nonumber \\
\Gamma^{pp, \, p_2, \, p_3}_{\tau \tau'; \tau_{N} {\tau'}_{N}}(\v{k},k_F,\theta) & =
& \Gamma^{pp, \, p_1, \, p_2}_{\tau \tau'; \tau_{N} {\tau'}_{N}}(\v{k},k_F,\theta)| \\
&&\delta(\cos\theta-(\v{k-q})_{z}/|\v{k-q}|) \rightarrow
\delta(\cos\theta-(\v{h'+t})_{z}/|\v{h'+t}|)\, . \nonumber
\end{eqnarray}
The next step is to implement the isospin--summation to obtain:
\begin{eqnarray}
\label{gamma2nppp}
\Gamma^{pp, \, p_1}_{np} & = & 4 (
{\Gamma}^{pp, \, p_1}_{11,11} + {\Gamma}^{pp, \, p_1}_{00,00} + 2 {\Gamma}^{pp, \, p_1}_{11,01})\, ,\\
\Gamma^{pp, \, p_2}_{np}  & = &
5 {\Gamma}^{pp, \, p_2}_{11,11} + {\Gamma}^{pp, \, p_2}_{00,00} +  {\Gamma}^{pp, \, p_2}_{11,00} +
5 {\Gamma}^{pp, \, p_2}_{00,11} \nonumber \\
&& -2 {\Gamma}^{pp, \, p_2}_{11,01} + 6 {\Gamma}^{pp, \, p_2}_{01,11} - 2 {\Gamma}^{pp, \, p_2}_{00,01} -
2 {\Gamma}^{pp, \, p_2}_{01,00} + 4 {\Gamma}^{pp, \, p_2}_{01,01}\, , \nonumber \\
\Gamma^{pp, \, p_3}_{np}  & = &
5 {\Gamma}^{pp, \, p_3}_{11,11} + {\Gamma}^{pp, \, p_3}_{00,00} + {\Gamma}^{pp, \, p_3}_{11,00} +
5 {\Gamma}^{pp, \, p_3}_{00,11} \nonumber \\
&& - 2 {\Gamma}^{pp, \, p_3}_{11,01} - 6 {\Gamma}^{pp, \, p_3}_{01,11} - 2 {\Gamma}^{pp, \, p_3}_{00,01} +
2 {\Gamma}^{pp, \, p_3}_{01,00} - 4 {\Gamma}^{pp, \, p_3}_{01,01}\, , \nonumber
\end{eqnarray}
where the $(\v{k},k_F,\theta)$--dependence of all these functions has
been omitted for simplicity.
In a similar way, for the $\vec{\Lambda} pp  \rightarrow  npp$ reaction we have:
\begin{eqnarray}
\label{gamma2hh}
\Gamma^{pp, \, p_1, \, p_2}_{pp} & = &
16 {\Gamma}^{pp, \, p_1, \, p_2}_{11,11}\, , \\
\Gamma^{pp, \, p_1, \, p_3}_{pp} & = & 4 (
{\Gamma}^{pp, \, p_1, \, p_3}_{11,11} + {\Gamma}^{pp, \, p_1, \, p_3}_{11,00} -
2 {\Gamma}^{pp, \, p_1, \, p_3}_{11,01})\, , \nonumber \\
\Gamma^{pp, \, p_2, \, p_3}_{pp} & = &
{\Gamma}^{pp, \, p_2, \, p_3}_{11,11} + {\Gamma}^{pp, \, p_2, \, p_3}_{00,00} +
{\Gamma}^{pp, \, p_2, \, p_3}_{00,11} +
2 {\Gamma}^{pp, \, p_2, \, p_3}_{11,01} - 2 {\Gamma}^{pp, \, p_2, \, p_3}_{01,11} \nonumber \\
&& +2 {\Gamma}^{pp, \, p_2, \, p_3}_{00,01} - 2 {\Gamma}^{pp, \, p_2, \, p_3}_{01,00}
- 4 {\Gamma}^{pp, \, p_2, \, p_3}_{01,01}\, . \nonumber
\end{eqnarray}
The final point is to employ Eq.~(\ref{decwpar3}) to implement the local density approximation.
We have, then:
\begin{eqnarray}
\label{g2pp}
\Gamma^{pp}_{np}(\theta) & \equiv & \Gamma^{pp, \, p_1}_{np}(\theta) +
\Gamma^{pp, \, p_2}_{np}(\theta) +\Gamma^{pp, \, p_3}_{np}(\theta)\, , \\
\Gamma^{pp}_{pp}(\theta) & \equiv & \Gamma^{pp, \, p_1, \, p_2}_{pp}(\theta) +
\Gamma^{pp, \, p_1, \, p_3}_{pp}(\theta) + \Gamma^{pp, \, p_2, \, p_3}_{pp}(\theta)\, . \nonumber
\end{eqnarray}
Finally, the $pp$ contribution to $N_p^{2N}(\theta)$ is obtained by Eq.~(\ref{np1f}).

We then consider the $2N$ decay contribution from the $ph$ diagram of Fig.~\ref{2Nfsi}.
We follow the same steps of the former contributions. We start by
introducing the partial, isospin--dependent decay widths for $\vec \Lambda np\to nnp$:
\begin{eqnarray}
\label{gamphp1}
\Gamma^{ph, \, p_1}_{\tau \tau'; \tau_{N} {\tau'}_{N}}(\v{k},k_F,\theta) & =
& \frac{(G_F m_{\pi}^2)^{2}}{(2 \pi)^{5}}
\left(\frac{f_{\pi}^2}{4 \pi}\right)^{2}  \frac{1}{m_{\pi}^4}
\frac{2}{(2 \pi)^2} \;
\int d  \v{q}
\int d  \v{t} \int d \v{h} \int d \v{h'} \,
\theta(q_{0}) \\
&& \times \theta(|\v{k}-\v{q}| - k_F)
\theta(|\v{h}-\v{t}| - k_F) \, \theta(|\v{h}-\v{t}+\v{q}| - k_F)
\theta(k_F - |\v{h}|)
\nonumber \\
&&
\times \theta(k_F - |\v{h}+\v{q}|) \, \theta(|\v{h'}+\v{t}| - k_F) \,
\theta(k_F - |\v{h'}|) \; \delta(\cos\theta-(\v{k-q})_{z}/|\v{k-q}|)
\nonumber \\
&&
\times \delta(q_0 - (E_N(\v{h}-\v{t}+\v{q}) - E_N( \v{h})+E_N(\v{h'}+\v{t})-E_N(\v{h'}))
\nonumber \\
& &
\times \frac{{\cal S}^{ph}_{\tau \tau'; \tau_{N} {\tau'}_{N}}(q,t)}
{E_N(\v{h}-\v{t})-E_N(\v{h})+E_N(\v{h'}+\v{t})-E_N(\v{h'})}
\nonumber \\
& &
\times \frac{1}
{E_N(\v{h}+\v{q}-\v{t})-E_N(\v{h}+\v{q})+E_N(\v{h'}+\v{t})-E_N(\v{h'})}\, , \nonumber \\
&& \nonumber \\
\Gamma^{ph, \, p_2}_{\tau \tau'; \tau_{N} {\tau'}_{N}}(\v{k},k_F,\theta) & =
& \Gamma^{ph, \, p_1}_{\tau \tau'; \tau_{N} {\tau'}_{N}}(\v{k},k_F,\theta)| \\
&&\delta(\cos\theta-(\v{k-q})_{z}/|\v{k-q}|) \rightarrow
\delta(\cos\theta-(\v{h-t+q})_{z}/|\v{h-t+q}|)\, ,\nonumber \\
&& \nonumber \\
\Gamma^{ph, \, p_3}_{\tau \tau'; \tau_{N} {\tau'}_{N}}(\v{k},k_F,\theta) & =
& \Gamma^{ph, \, p_1}_{\tau \tau'; \tau_{N} {\tau'}_{N}}(\v{k},k_F,\theta)| \\
&&\delta(\cos\theta-(\v{k-q})_{z}/|\v{k-q}|) \rightarrow
\delta(\cos\theta-(\v{h'+t})_{z}/|\v{h'+t}|)\, , \nonumber
\end{eqnarray}
where $p_1$, $p_2$ and $p_3$ indicate the position of the emitted proton. In a similar way,
for the reaction $\vec{\Lambda} pp  \rightarrow  npp$ we have:
\begin{eqnarray}
\label{gamphp1p3}
\Gamma^{ph, \, p_1, \, p_3}_{\tau \tau'; \tau_{N} {\tau'}_{N}}(\v{k},k_F,\theta) & =
& \frac{(G_F m_{\pi}^2)^{2}}{(2 \pi)^{5}}
\left(\frac{f_{\pi}^2}{4 \pi}\right)^{2}  \frac{1}{m_{\pi}^4}
\frac{2}{(2 \pi)^2} \;
\int d  \v{q}
\int d  \v{t} \int d \v{h} \int d \v{h'} \,
\theta(q_{0}) \\
&& \times \theta(|\v{k}-\v{q}| - k_F)
\theta(|\v{h}-\v{t}| - k_F) \, \theta(|\v{h}-\v{t}+\v{q}| - k_F)
\theta(k_F - |\v{h}|)
\nonumber \\
&&
\times \theta(|\v{h'}+\v{t}| - k_F) \,
\theta(k_F - |\v{h'}|) \nonumber \\
&&\times
(\delta(\cos\theta-(\v{k-q})_{z}/|\v{k-q}|)+\delta(\cos\theta-(\v{h'+t})_{z}/|\v
{ h'+t } |))/2
\nonumber \\
&&
\times\delta(q_0 - (E_N(\v{h}-\v{t}+\v{q}) - E_N( \v{h})+E_N(\v{h'}+\v{t})-E_N(\v{h'}))
\nonumber \\
& &
\times\frac{{\cal S}^{ph}_{\tau \tau'; \tau_{N} {\tau'}_{N}}(q,t) }
{(E_N(\v{h}-\v{t})-E_N(\v{h})+E_N(\v{h'}+\v{t})-E_N(\v{h'}))^{2}}\, ,\nonumber \\
&& \nonumber \\
\Gamma^{ph, \, p_2, \, p_3}_{\tau \tau'; \tau_{N} {\tau'}_{N}}(\v{k},k_F,\theta) & =
& \Gamma^{ph, \, p_1, \, p_3}_{\tau \tau'; \tau_{N} {\tau'}_{N}}(\v{k},k_F,\theta)| \\
&&\delta(\cos\theta-(\v{k-q})_{z}/|\v{k-q}|) \rightarrow
\delta(\cos\theta-(\v{h-t+q})_{z}/|\v{h-t+q}|)\, , \nonumber
\end{eqnarray}
The next step is to implement the isospin summation to obtain:
\begin{eqnarray}
\label{gamma2np_ph}
\Gamma^{ph, \, p_1}_{np} & = & 4 (
-{\Gamma}^{ph, \, p_1}_{11,11} + {\Gamma}^{ph, \, p_1}_{11,00} + 2 {\Gamma}^{ph, \, p_1}_{11,01})\, , \\
\Gamma^{ph, \, p_2}_{np}  & = &
-3 {\Gamma}^{ph, \, p_2}_{11,11} + {\Gamma}^{ph, \, p_2}_{00,00} + 5 {\Gamma}^{ph, \, p_2}_{00,11} +
{\Gamma}^{ph, \, p_2}_{11,00} \nonumber \\
&& + 6 {\Gamma}^{ph, \, p_2}_{01,11} -2 {\Gamma}^{ph, \, p_2}_{01,00} - 2 {\Gamma}^{ph, \, p_2}_{11,01} -
2 {\Gamma}^{ph, \, p_2}_{00,01} +  4 {\Gamma}^{ph, \, p_2}_{01,01}\, ,  \nonumber \\
\Gamma^{ph, \, p_3}_{np} & = &
-3 {\Gamma}^{ph, \, p_3}_{11,11} + {\Gamma}^{ph, \, p_3}_{00,00} + 5 {\Gamma}^{ph, \, p_3}_{00,11} +
 {\Gamma}^{ph, \, p_3}_{11,00} \nonumber \\
&& -6 {\Gamma}^{ph, \, p_3}_{01,11} + 2 {\Gamma}^{ph, \, p_3}_{01,00} - 2 {\Gamma}^{ph, \, p_3}_{11,01} -
2 {\Gamma}^{ph, \, p_3}_{00,01} - 4 {\Gamma}^{ph, \, p_3}_{01,01}\, , \nonumber
\end{eqnarray}
where the $(\v{k},k_F,\theta)$--dependence of all these functions has
been omitted for simplicity.
In a similar way, for the $\vec{\Lambda} pp  \rightarrow  npp$ reaction we have:
\begin{eqnarray}
\label{gamma2pp_ph}
\Gamma^{ph, \, p_1, \, p_3}_{pp} & = & 4 (
-{\Gamma}^{ph, \, p_1, \, p_3}_{11,11} + {\Gamma}^{ph, \, p_1, \, p_3}_{11,00} -
2 {\Gamma}^{ph, \, p_1, \, p_3}_{11,01})\, , \\
\Gamma^{ph, \, p_2, \, p_3}_{pp}  & = &
{\Gamma}^{ph, \, p_2, \, p_3}_{11,11} + {\Gamma}^{ph, \, p_2, \, p_3}_{00,00}
+ {\Gamma}^{ph, \, p_2, \, p_3}_{00,11} +
{\Gamma}^{ph, \, p_2, \, p_3}_{11,00} - 2 {\Gamma}^{ph, \, p_2, \, p_3}_{01,11} \nonumber \\
&& -2 {\Gamma}^{ph, \, p_2, \, p_3}_{01,00} + 2 {\Gamma}^{ph, \, p_2, \, p_3}_{11,01} +
2 {\Gamma}^{ph, \, p_2, \, p_3}_{00,01} - 4 {\Gamma}^{ph, \, p_2, \, p_3}_{01,01}\, , \nonumber
\end{eqnarray}
One thus has to perform the local density approximation, through Eq.~(\ref{decwpar3}), to obtain:
\begin{eqnarray}
\label{g2ph}
\Gamma^{ph}_{np}(\theta) & \equiv & \Gamma^{ph, \, p_1}_{np}(\theta) +
\Gamma^{ph, \, p_2}_{np}(\theta) +\Gamma^{ph, \, p_3}_{np}(\theta)\, ,\\
\Gamma^{ph}_{pp}(\theta) & \equiv & \Gamma^{ph, \, p_1, \, p_3}_{pp}(\theta) +
\Gamma^{ph, \, p_2, \, p_3}_{pp}(\theta)\, , \nonumber
\end{eqnarray}
and finally the $ph$ contribution to $N_p^{2N}(\theta)$ is obtained by Eq.~(\ref{np1f}).


\end{document}